# Threat Modelling in Virtual Assistant Hub Devices Compared With User Risk Perceptions

Beckett LeClair

[160064531]

## Abstract

Despite increasing uptake, there are still many concerns as to the security of virtual assistant hubs (such as Google Nest and Amazon Alexa) in the home. Consumer fears have been somewhat exacerbated by widely-publicised privacy breaches, and the continued prevalence of high-profile attacks targeting IoT networks. Literature suggests a considerable knowledge gap between consumer understanding and the actual threat environment; furthermore, little work has been done to compare which threat modelling approach(es) would be most appropriate for these devices, in order to elucidate the threats which can then be communicated to consumers. There is therefore an opportunity to explore different threat modelling methodologies as applied to this context, and then use the findings to prototype a software aimed at educating consumers in an accessible manner. Five approaches (STRIDE, CVSS, Attack Trees (a.k.a. Threat Trees), LINDUNN GO, and Quantitative TMM) were compared as these were determined to be either the most prominent or potentially applicable to an IoT context. The key findings suggest that a combination of STRIDE and LINDUNN GO is optimal for elucidating threats under the pressures of a tight industry deadline cycle (with potential for elements of CVSS depending on time constraints), and that the trialled software prototype was effective at engaging consumers and educating about device security. Such findings are useful for IoT device manufacturers seeking to optimally model threats, or other stakeholders seeking ways to increase information security knowledge among consumers.

## Acknowledgements

Over the course of the project, I have received valuable assistance and support from many. I would like to thank Dr Kaniz Fatema for her work as project supervisor, as well as Dr Sylvia Wong for her insights as course director for MSc Computer Science. Furthermore, I would like to thank all those who took part in the primary research. I would finally like to express my gratitude to those who have supported me through my studies, especially my fiancé and also my mother, without whose encouragement I would not have been able to pursue my academic goals.

# List of Contents

# 1. Introduction

*1.1 General Introduction*

Threat modelling is defined as a collection of activities for understanding and improving security by identifying potential problems and brainstorming ways to mitigate them (OWASP, 2021). The key benefit is that it allows for predicting and mitigating threats before they arise, avoiding the often large opportunity cost of post-disaster cleanup. There are a great many methods of threat modelling currently in use; this paper will primarily focus on five: Microsoft's STRIDE, attack trees (sometimes known as threat trees), CVSS, LINDUNN GO, and Quantitative TMM.

The Internet of Things (IoT) refers to the network of physical objects with embedded, internet-connected systems, such as 'smart logistics' devices and the rise of the 'smart home' consumer products (Oracle, 2021). One such smart home device will be the focus of this work, that being the virtual assistant hub devices. These voice-controlled products are capable of speaking back to the user, executing 'apps' or 'skills', and communicating with a variety of other devices both on the home network and beyond. Examples of such devices include Amazon Echo (Alexa) and Google Home. Such devices are still in their relative infancy but are increasingly being installed in homes across the world. Despite this, there are growing concerns about the safety of such products; for example, there have been cases of unintended data transfer of private conversations, sparking discussion about threats to privacy and confidentiality (Hern, 2019). These concerns, combined with the expanding influence of IoT networks and the past attacks on them (such as Mirai, Silex and BrickerBot) ultimately point to a need to examine the threat environments of virtual assistant hub devices as a whole.

The purpose of this study is to compare popular threat modelling approaches by applying them to the context of the virtual assistant hub device, and gauging their effectiveness by a number of attributes. Following on from this, we will investigate how this information can be used to improve consumer awareness and understanding of threats to their home IoT networks. Much work has been done in threat modelling comparisons for various other IoT devices, from cars to healthcare systems, but there is currently a gap in the literature for the virtual assistant hub.

The main questions to be answered are as follows:

1.  What is the most effective way to model threats associated with a virtual assistant hub in the home? We will answer this by conducting a literature review of existing model theory and recent developments in IoT security, creating a matrix of qualities by which to rank threat models, conducting modelling activities according to a selection of approaches, and finally plotting the outcomes against the matrix.

2.  What is the difference between the detailed threat environment and the environment which is perceived by the user? This will be answered by investigating literature concerning consumer attitudes to home IoT devices and questioning a test group, then comparing this with the threats we have uncovered from our threat modelling activities.

3.  How may we bridge the gap between consumer knowledge and best practice? To answer this final question, we will create a small software prototype with the aim of educating users using aspects of what we determine to be the optimal threat models for this context. The software, while not the main deliverable of this project (that status belonging to the threat modelling analysis and conclusions), will be a useful exercise for testing the efficacy of a novel program designed to narrow the consumer knowledge gap.

*1.2 Problem Description*

Virtual assistant hub devices continue to increase in uptake among consumers, with a business model that prioritises fast product development and release. This consequently results in lesser emphasis being placed upon aspects of development such as threat modelling. There is a distinct lack of literature regarding which threat modelling techniques are optimal for this device scenario, and IoT home devices in the broader sense; this is our primary problem.

Furthermore, research into user perceptions of threats indicates significant gaps in user understanding which may lead to negligent or otherwise dangerous behaviour. There also seems to be a significant amount of distrust of these devices, somewhat paradoxically when one considers their increasing uptake among consumers. Both of these findings may be remedied by determining a way to use threat modelling to educate consumers (both in terms of how to secure their home IoT networks and what they should know before purchasing one), thus increasing the general device security of the general consumer population. The main issue here is finding out how we can bridge the knowledge gaps in a manner which is accessible to the layperson, a question which we can only answer after determining the optimal threat modelling approaches.

# 2. Context and Literature Review

*2.1 Threat Modelling*

Today's consumer products are becoming increasingly integrated with computing, and high competition means new products are released quickly and regularly. With the emphasis on speed of development, it is becoming ever-more imperative to undertake threat modelling activities early on, so as to avoid the high cost (in terms of both expenditure and missed sales revenue) of fixing later in the life cycle.

One of the most important works undertaken into comparing threat modelling tools is that by Shevchenko et al.. A total of twelve are compared and contrasted, making for a comprehensive overview of available methodologies. This, combined with the shared amount of expertise between the authors, makes this white paper one of our 'best' sources. It is, however, potentially becoming outdated; being from August 2018 makes it relatively modern but the rate of technological and theoretical development continues to expand at a rapid rate. Furthermore, while it compares the general attributes of the different models, it is not assessing their applicability to a particular context such as ours. It is, however, a very detailed and credible starting point.

Microsoft's STRIDE (standing for each of its categories - Spoofing, Tampering, Repudiation, Information Disclosure, Denial of Service, and Escalation of Privilege) is considered one of the most 'mature' of the techniques, making it a dominant force, though it comes with several significant drawbacks (Shevchenko et al., 2018). Shull agrees with this former notion, regarding it as 'state of the practice'. Drawbacks include the fact it is generally very time consuming (a claim we investigate later in this paper). It is also increasingly complex to apply as the modelled system becomes more intricate, meaning that a high level of expertise is required to make the process accurate and worthwhile. However, even if the analysis is performed by top experts, there is still a 'moderately high rate of false negatives' (Shevchenko et al., 2018). This means that it is generally insufficient alone, and further time must be spent supplementing it with other threat modelling techniques, making it a generally large investment. When we take this and the high knowledge barrier into account, it may not be an optimal approach for making threat modelling accessible to the average virtual assistant hub consumer, though we should investigate this later to be certain. Despite this, it is regularly applied in industry to both physical systems and non-physical, suggesting that it has at least some merit for IoT applications, making it of interest to this study. For example, it has been successfully applied in practice to telehealth systems (Abomhara et al., 2015). Other experts in industry generally attest to its usefulness; for example, Shull regards it as providing 'the greatest variability' of modern approaches (Shull, 2016). As we will cover shortly, however, there may be a limit to how far this 'variability' can get us.

CVSS (the Common Vulnerability Scoring System) is another tool which is often used in threat modelling. Developed by NIST and owned by FIRST, it allows for relative scoring of severity levels in determined threats (FIRST, 2019). However, there is some concern as to the lack of algorithm transparency and subjectivity of results, with different experts able to get different scores for the same scenarios (Shevchenko et al., 2018). Others with experience in the field of cyber security have considered prior versions of CVSS to have some issues with the weight given to different characteristics of the score, suggesting that tweaks would make it more accurate and therefore more worthwhile (Scarfone & Mell, 2009). However, this source is quite old and the CVSS scoring system has since been updated, limiting the impact of its claim to a modern reader. It is another tool which is still popular today in modelling both cyber and cyber-physical threat environments, suggesting that it is another tool of some use to IoT contexts - therefore we shall come back to it later.

Attack trees are one of the oldest threat modelling tools still used today, owing to their versatility and ease of understanding. However, they are only really effective when created by experts in a particular application (Shevchenko et al., 2018). This could potentially make them less useful to us, when we consider our aim of using threat modelling approaches to educate those with little to no background in cybersecurity. Literature support for attack trees extends far back in the literature with multiple analysts and proponents (Amoroso, 1994) (Saltner et al., 1998). These older works are limited in their applicability to a modern context by virtue of their age, but the longstanding nature of the technique means they may potentially be of use to us, and thus we will explore them again later.

Quantitative TMM is a combination method consisting of STRIDE, CVSS, and attack tree activities. It's modernity makes it potentially applicable to modern contexts such as IoT applications, and therefore it is of interest to us. Furthermore, there are examples where it has been used effectively as a modelling approach for cyber-physical applications (Aufner, 2019). Aufner's study is particularly important as it is relatively recent and shows direct relevance of the technique to the IoT context which we are considering. It is intended by its creators to be 'asset-centric' for ease of business use, and to provide a 'definitive, scientific approach' which is versatile across a wide range of scenarios (Potteiger et al., 2016). Interestingly, this is at odds with multiple other expert opinions who argue that there is no method that truly suits every scenario well, as we will discuss below. This claim is perhaps not as strong as others, though, considering it comes from the creators of the approach. It is also important to note, when one considers the complexity of each individual component alone, it is likely that this method is extremely time consuming and requires a lot of effort and expert knowledge.

The base framework of LINDUNN is considered useful for finding out relevant mitigation steps for threats, despite it being potentially labour-intensive (Shevchenko et al., 2018). It is not as well-represented in the literature as other approaches, suggesting that it is not as widely used. It's most recent offshoot, LINDUNN GO, aims to be more lightweight and accessible to those with less cybersecurity (DistriNet Research Group, 2021). However, as there is little discussion about this approach and the only source for these claims so far is the approach's own documentation, we cannot put much weight in this statement without investigating for ourselves.

Delving further into the idea of specific approaches being applicable to IoT contexts, Nurse et al. propose that current methodologies are 'inadequate' for four main reasons - the basis of 'one time' assessment in a constantly evolving landscape, changing system boundaries contrasted with limited knowledge, difficulty understanding the individual connections (termed 'glue' in their paper), and failure to consider assets a valid attack platform (Nurse et al., 2017). Despite the fact that this paper is a few years old, it is useful in that it relates explicitly to the kinds of contexts we are exploring. Thus, we should value it with some weight, and assume it may be the case that no single method we will investigate will perform well on all bases.

Overall, there are a variety of techniques at experts' disposal, but opinions differ as to which are the most applicable, with multiple authors rejecting the idea of a 'one size fits all' threat modelling tool. Shull argued that no technique will fit every application with equal efficacy, with 'substantial trade-offs' in terms of outcomes (Shull, 2016). This is reiterated by Aufner, who suggests a 'gap between threat modelling frameworks and IoT', perhaps lending to the argument that IoT is a more modern development that requires a more modern way of approaching the situation (Aufner, 2019). Additionally, Nurse et al. lend weight to this. Thus, we will need to conduct independent work to assess which is most applicable to our virtual assistant hub context. Trends tend toward encouraging STRIDE and CVSS-based approaches, though whether this is due to their genuine applicability to the modern context or simply their longstanding reputation has yet to be determined. Additionally, there is little evidence for academic exploration or evaluation of LINDUNN GO, presenting an interesting opportunity for this study.

*2.2 Device Context*

General consensus among available sources states that there are significant, as-yet unaccounted for security issues with IoT devices as a whole. Due to the speed of market development and the infancy of the technologies involved, products often reach consumers 'without proper safeguards in place' to ensure sufficient safety, with a trend that 'security tends to lag behind innovation' (Rizvi et al., 2020). That these quotations come from a recent study in a well-regarded IoT journal tells us that the problem is still rife and relevant. The sentiment is echoed throughout multiple works; for example, Pacheco and Hariri state that current methodologies are 'far from satisfactory' when it comes to facing modern, complex threats (though it must be pointed out that this paper is older, limiting the strength of its claims) (Pacheco & Hariri, 2016). As discussed earlier, Aufner also found a 'gap' between current applications of threat modelling and the variety of IoT systems and components (Aufner, 2019).

'Hub' style devices like the virtual assistant hub make for attractive targets for threat actors due to their centrality on the network; they can often be manipulated to compromise the entire network itself (Rizvi et al., 2020). Due to this, we believe that work must be done imminently to promote better security knowledge among the users of such devices. Supporting this notion of a highly attractive target, Bugeja et al. define six key categories of potential threat actors for these devices - nation states, terrorists, organised criminals, hacktivists, thieves, and lone hackers (Bugeja et al., 2017). User perceptions place some emphasis of the hypothetical threat of nation state surveillance, as we will cover later. A pivotal work by Cho et al. uncovered sixteen different types of threats in voice assistant applications, which often function as the 'core' of hub devices - these threats ranged from simple eavesdropping to total device compromise through privilege escalation (Cho et al., 2018). Lit et al. also found six separate attack surfaces in Amazon's Alexa, lending weight to this (Lit et al., 2021). It should also be noted that Lit's work is very recent, suggesting that the vulnerability problem has yet to be solved. Perhaps exacerbating the weakness in these devices is the purportion by some academics that default-configured hub devices are 'too weak' to provide sufficient protection for users (Seeam et al., 2019). That users tend to lack knowledge on how to effectively configure devices (as we will see later) only worsens the problem.

A particular threat vector which is easily exploited is the app or 'skill' function present on these kinds of devices, as malicious payloads can easily be stored within source code and then executed when an unknowing user installs and runs the function - though other key threats include MITM (man in the middle) and denial of service attacks (Rizvi et al., 2020). In a key study by Hu et al., it was found that the security vetting system for downloadable apps on these kinds of devices are easily fooled, or the systems are lacking in the first place, in an experiment where intentionally exploitable apps were placed on devices (Hu et al., 2020). When one considers that such skills are often used to perform highly sensitive tasks such as sending money or controlling home security devices, this becomes a key concern, and highlights the need for work to be done quickly in educating consumers of potential dangers.

Further evidence of threats includes the use of machine learning in one study to classify encrypted traffic (Jackson & Camp, 2018). This could, for example, be used to classify signals from connected devices, allowing the attacker to gauge a pattern of when the user may not be home. The threat of issuing commands without the user's consent (as in the case of dolphin attacks) is also an issue, with Lei et al. finding that 'no access control is deployed' in Alexa devices, since 'vendors consider that all the voice commands from the Alexa service are benign' (Lei et al. 2018). This concept of sound-based attacks was also demonstrated in practice by other studies (Haack et al., 2017).

Additionally, threat analyses of other IoT devices show similar susceptibilities, perhaps suggesting that these problems are intrinsic to the IoT itself as well as an element of it being due to the nature of the individual device. An analysis of smart camera technology revealed various vulnerabilities, allowing for exploitations such as eavesdropping, MITM attacks, and denial of service (Alhabi & Aspinall, 2018). Multiple analyses of the 'smart home' as a whole also produced a wide range of security problems (Zeng et al., 2017) (Kavallieratos et al., 2019).

*2.3 Consumer Understanding*

Despite what we have learned about the range of vulnerabilities and difficulties in threat modelling, security is still at the forefront of user concerns when they buy and install the product in their homes. It is important to note before continuing that the studies to be discussed in this section focus on consumers from the United States, potentially limiting their applicability to UK consumers. Thus, part of our work will involve collecting feedback from UK consumers in order to validate or disprove any claims.

Zeng et al. performed a study which uncovered some interesting conclusions on how end users approach security with their IoT devices - it is important to note, however, that the sample size for this study was relatively small (15 individuals) and respondent demographics were somewhat biased towards males. However, the information can still be useful to us as it sheds some light on individual insights and potential trends, even if they cannot be confidently generalised to the entire population.

Participants were mainly concerned about the threat of eavesdropping, and were most concerned about the possibility of the device manufacturers themselves listening in on conversations they had around the devices, though a few also expressed concern about their country's government listening in for law enforcement purposes or political gain (Zeng et al., 2017). An excerpt of a response illustrating this can be seen in Figure 1.

> **"I am beefing up on operational security in a big way, because I have spoke publicly against fascism, and I work in a publicly funded institution, I expected to be targeted at some point."**
> - *Respondent (Zeng et al, 2017)*

Figure 1: Evidence of eavesdropping concerns among end users of IoT home devices

Responses like these are interesting as they suggest a dissonance between consumer's thoughts and their actions when it comes to purchasing these devices. Other key conclusions from this study are that consumers' understanding of network threats is 'sparse', and a proposal that external guidance may be required in order to better educate these users, with the responsibility falling upon those who already have some IoT security expertise (Zeng et al., 2017). This proposal in particular is something we will test the efficacy of later via a novel software approach.

A study that also lends weight to this proposal looked at manuals and user guides for 270 IoT devices and concluded that manufacturers give users 'too little information about the security features of their devices' (Blythe et al., 2019). This paper provides necessary support as it covers a much more extensive sample pool to reach a similar sentiment.

Another study focused on consumer opinions with a slightly larger sample size (and therefore potentially more weight to its arguments) was conducted by Harney et al.. Figure 2 shows evidence of participants showing similar focus on fears of eavesdropping.

> **"My husband is paranoid Google's listening to him about conversations at work."**
>
> **"Just from a general big brother perspective, I think you're naive to think that we're not being watched…"**
>
> **"You have no idea when it's communicating to the manufacturer or what it's communicating to the manufacturer."**
>
> - Respondents (Harney et al, 2021)

Figure 2: Sentiments about eavesdropping and data collection among users

This further supports the idea that actor mistrust and concerns about confidentiality are primary concerns among users. However, some respondents also did not care or felt the risks of being personally targeted were rather small, something which was also hinted at in the previous consumer study (Harney et al., 2021) (Zeng et al., 2017). Interestingly, users felt the primary responsibility for security fell to both the individual and the manufacturer, with privacy considerations themselves falling more heavily to the manufacturer alone (Harney et al., 2021). This suggests that users may be open to taking on greater responsibility by learning about aspects of device security, something which we will test later.

*2.4 Past Attacks*

The prevalence of past attacks and accidents plays significantly into the context of the problem, especially with regard to consumer attitudes to security.

Attacks such as the Mirai botnet, Silex, Stuxnet, and BrickerBot have been targeted at IoT devices in the past, with often devastating monetary, safety, and privacy consequences. A well-publicised case of uncovered security concerns took place when Ring doorbells were shown to have a vulnerability allowing potential hackers to view security footage from user homes (O'Donnell, 2019). Even when all data transmissions in an IoT device are encrypted, there is still the potential for someone sniffing the network to determine a usage pattern (Jackson & Camp, 2018). These patterns may then, for example, be used to determine times when the user is not usually at home, which has obvious implications for physical crime. Another experiment showed that Alexa devices in particular may have the security PIN brute-forced under certain circumstances (Haack et al., 2021).

On top of the threats posed by bad actors, the nature of the devices also allows for accidental breaches of intended use to occur, with dangerous or expensive consequences. For example, in one case a child speaking with an Amazon Alexa device caused an unintended purchase of cookies and a $160 dollhouse (Chung et al., 2017). This was further confirmed as a possible vector for intentional exploitation in a study of both Alexa and Google Home devices (Lei et al., 2018).

In a similar, popularized case, a private conversation was accidentally broadcast over the internet after the Alexa hub reportedly misinterpreted parts of the conversation as command words (Griffin, 2018).

Something which may fuel the eavesdropping concerns of many users is the volume of data collection and processing which occurs in these devices. For example, Amazon uses human employees to listen to saved recordings of users interacting with their devices for the purposes of 'improvement' to the voice processing software (Lit et al., 2021). The same is true of Google devices (Verheyden et al., 2019). Flikkema and Cambou posed some scenarios at an IEEE summit about the potential for these kinds of features to be misused, as can be shown in the excerpt contained in Figure 3 (Flikkema & Cambou, 2017).

> **"You are recovering from a cold and still coughing. Your home digital assistant's skill of recognizing coughing captures this in its database and is later harvested by your health insurance provider. Your health insurance premiums inexplicably go up, and there is no way to trace why. At a group conversation in your home, someone makes a sarcastic joke about a government, either domestic or foreign, that includes certain key words on a watch list. The next evening, law enforcement officials ring your doorbell."**
> - Excerpt (Flikkema & Cambou, 2017)

Figure 3: Expanding on the potential for misuse via 'service improvement' processing

When one considers the previously discussed opportunity for recordings to be made via misinterpreted commands, this opens up the opportunity for unintended audio data to be heard by manufacturer employees. This indeed proved to be the case when Apple employees reported having heard accidental recordings of private medical details, crimes, or users engaging in sexual intercourse; this clearly carries heavy implications for privacy and dignity (Hern, 2019).

All of this shows us that the technology is still currently insecure in many regards, and consumer fears of private information leakage may have some basis in reality. In the face of such vulnerabilities, a clear need arises for a way to educate users on the potential dangers and how they can make their devices safer.

# 3. Threat Modelling

## 3.1 STRIDE

To aid in all of the threat modelling approaches tested in this study, a context diagram (Figure 4) and DFD (Figure 5) were first created.

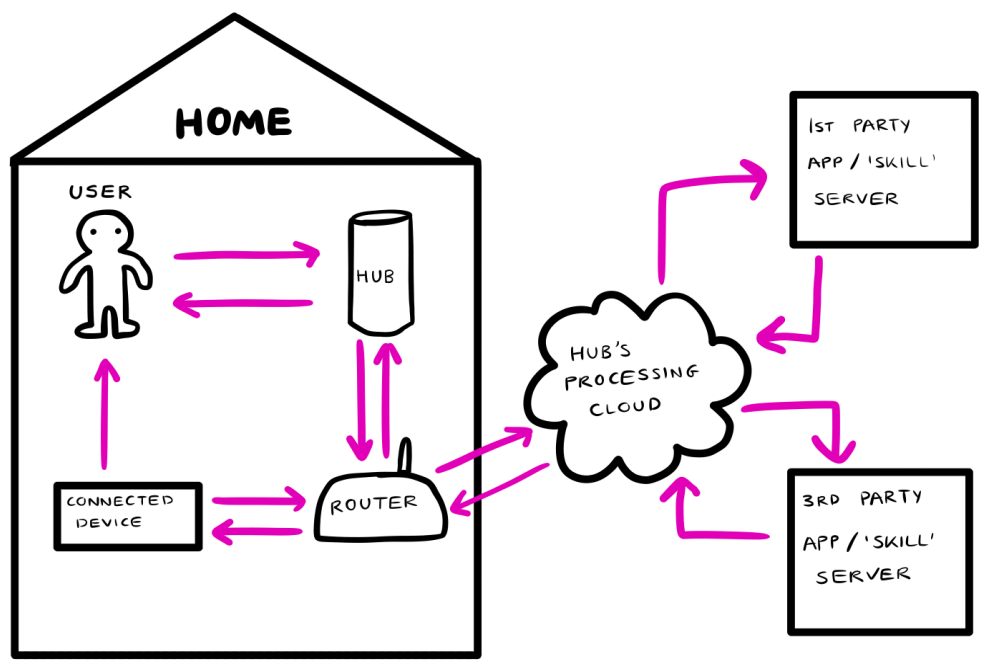

Figure 4: Context diagram for the typical virtual assistant hub in the home

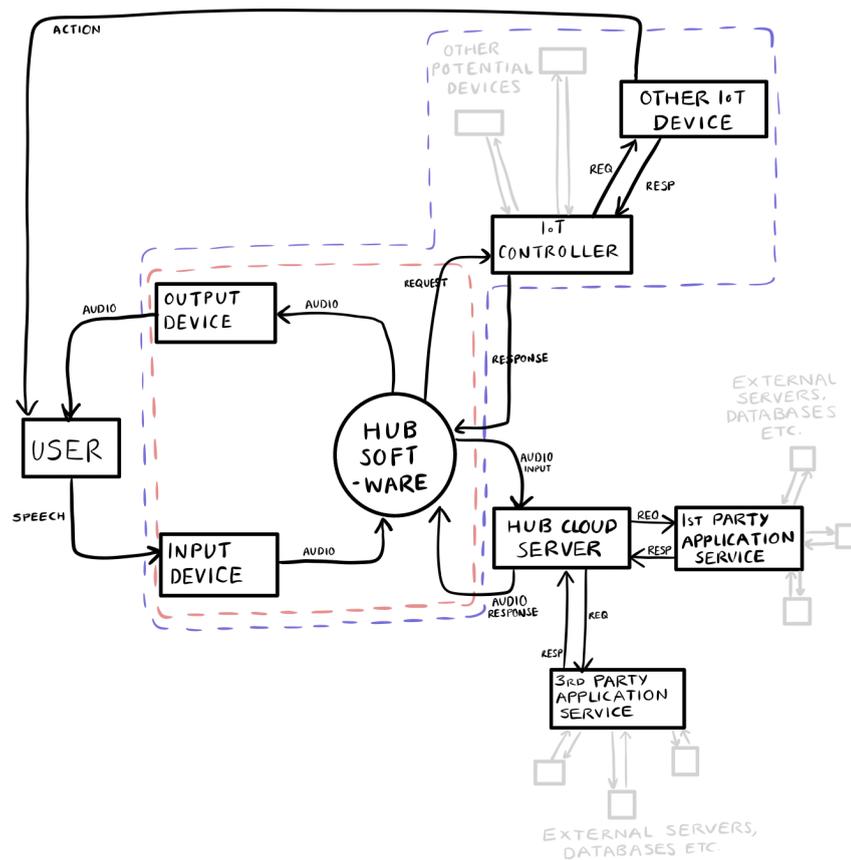

Figure 5: Data flow diagram for the typical virtual assistant hub in the home

As STRIDE is regarded as something of a de-facto industry standard for threat modelling, it would be useful to attempt to apply it to this context. Due to its popularity, some work has already been done in threat modelling IoT home devices using this framework. One study has carried out a STRIDE analysis of voice assistant applications as a whole, which covers both home digital assistants as well as mobile phone functions like the portable version of Apple's Siri (Cho et al., 2018). Though its scope is a little broader than the one we are considering here, much of the points they raise are still applicable, so their paper has been used as a starting point when producing the table below. Furthermore, research work has been undertaken with regard to conducting STRIDE analysis for the concept of the 'smart home' as whole (Kavallieratos et al., 2019). This paper also raises some useful concerns which have been further incorporated into this work.

Figure 6 below shows the elucidated threats from applying STRIDE, and Figure 7 shows the associated potential mitigations. Something to keep in mind is that we are threat modelling a type of device and not one specific model of device, so some of the identified threats and mitigations may seem broad, and may not apply to every single device of that type.

| | Potential Threats |
|---|---|
| **S** | 1. An actor may issue a voice command to another user's device from nearby, despite them not being the intended user. For example, one might say 'Alexa, add XYZ to my shopping basket'.<br><br>2. If the device lacks adequate measures for authentication, an attacker can spoof a legitimate signal from the device and transmit it to the device's processing server to execute a command.<br><br>3. If the device lacks adequate measures for authentication, an attacker can spoof a legitimate signal from the device and transmit it to a connected IoT device to execute a command or action. |
| **T** | 4. A legitimate voice command from the intended user is captured and modified by a 'man in the middle' in order to execute a malicious action.<br><br>5. If the device lacks sufficient checks for integrity of reply signals, the reply from the server to the device may be tampered by a 'man in the middle' in order to execute a malicious action.<br><br>6. If the device lacks adequate measures for authentication, requests by the device to other connected devices can be modified in a 'man in the middle' attack, similar to 4 above.<br><br>7. A connected device's legitimate 'action response' (for example, a lock signal that opens a door) is captured, modified, and then re-injected for execution. In our example this may, for example, put the door lock in an insecure state of operation. |
| **R** | *Repudiation threats are not very applicable to the context of this device. Repudiation-type threats are intrinsic to devices where there is some consequence to having carried out an action on a device, for example in a military setting or context where device usage may be used for blackmail. This argument is also supported by prior work (Cho et al., 2018).* |
| **I** | 8. The response signal from the server to the device could be sniffed for sensitive user information, such as a shipping address.<br><br>9. An action response from a connected device may be disclosed by someone sniffing the network. For example, this could be dangerous if an actor collects signals from an IoT door lock to determine a pattern for when the |
| | owner is not at home. |
| | 10. People in the vicinity may listen to commands the user issues to the device, and these may contain sensitive information. |
| | 11. Unintended user information or snippets of recorded voice commands may be heard by employees during routine analysis, or may be transferred to other parties through misinterpreted device commands, and these snippets may contain sensitive information. |
| **D** | 12. Interfering voice signals (including those produced by a 'dolphin attack') can be continuously injected to deny service to a legitimate voice command. |
| | 13. A malicious server reply may be injected which contains bad commands or data, congesting the network or cache. |
| | 14. A malicious action response may be injected with similar effects to 13 above. |
| **E** | 15. A malicious reply signal from the device server may be created using some known vulnerability (such as code injection, unpatched functions, or default admin passwords) to allow for privilege escalation. |
| | 16. A malicious action response signal from a connected device may be created using some known vulnerability (such as code injection, unpatched functions, or default admin passwords) to allow for privilege escalation. |

Figure 6 - STRIDE table of threats for the home virtual assistant device

| No | Potential Mitigation |
|---|---|
| 1 | The user can modify their device settings to require multiple levels of authentication (as in 2FA) before executing more powerful commands. Voice recognition settings may also be applied, though these are not foolproof and can also be awkward in some cases (for example, if the user has a cold and their voice is not recognised). |
| 2 | Data traffic must be checked for its authenticity at each stage of transport. Additionally, a strong encryption protocol must be used to transmit this data. |
| 3 | See 2 above. |
| 4 | Data traffic must be checked for integrity at each stage of transport. The system should keep a log of all failed checks and alert the user of each addition to this log. |

| 5 | See 4 above. |
|---|---|
| 6 | See 4 above. |
| 7 | See 4 above. |
| 8 | Sensitive information should never be transmitted using an unencrypted (or weakly-encrypted) channel. The device should always prompt the user to change passwords from default and set strong passwords. The user should not be allowed to select weak passwords or default ones. |
| 9 | See 8 above. Additionally, a setting could be included to generate fake traffic throughout times when the user is not at home, though this obviously presents a small cost for computing power. |
| 10 | Users must take care not to install their device in a place where eavesdropping is comparatively easy (for example, where walls or windows are thin). Users must constantly be aware of who may be listening when they issue a command. |
| 11 | Users can opt out of voice processing by humans (often under some 'data collection for improvement' setting) in most cases. Additionally, sensitive data should ideally be filtered out before audio is heard by employees. However, this system could never be foolproof, and to an extent this risk is unavoidable. |
| 12 | See 10 above. Additionally, users must be wary when they play media that may use the wake-up word for their device (for example, advertisements or films). To avoid this latter risk, there may be an option for a voice recognition approach, though this comes with drawbacks as assessed in 1 above. |
| 13 | See 4 above. Additionally, users should only install apps or 'skills' that they trust wholly, and only connect their device with other devices they trust wholly. |
| 14 | See 13 above. |
| 15 | See 4 above. Furthermore, the user must ensure their passwords are strong and their device is kept up to date with the latest secure patch level. |
| 16 | See 15 above. |

Figure 7: Potential mitigations for elucidated threats

From completing this activity, we can glean several pros and cons of the STRIDE methodology in the case of the home IoT context. First of all, the framework covers a wide range of threat types and therefore allows us to generate a great deal of novel ideas, giving us much to work with when it comes to predicting and mitigating threats.

Secondly, and perhaps most importantly, STRIDE tends to be quite standard practice and is thus easily understood and shared between experts. This means it is less time-consuming when trying to convey information between key stakeholders, something which is important when threats and their actors are ever-evolving. Had it not been for the ease of shared knowledge and prior work by previous authors, the task of completing the STRIDE analysis for this paper would have been far more difficult and laborious; the application of other authors' threat knowledge would have taken far longer due to essentially having to 're-code' it to fit the STRIDE format.

The final advantage of STRIDE in this context is that categorising the threats into the acronym's groups makes tackling them much more manageable - most of the threats within the same category have the same if not similar mitigations, making them efficiently handled in groups.

However, there are still drawbacks. Firstly, this approach does not assign relative importance to any of the threats, making it difficult to gauge which are most likely to be exploited, or which are likely to have the most impactful damage if they are. This makes it tricky to know which areas to prioritise, something which is especially detrimental if one considers the rapid development cycles characteristic of the home IoT industry. There is almost never time to cover every basis, so the most dangerous threats are the ones which are usually focused on. This relates closely to the second disadvantage - STRIDE threat modelling takes a long time to do well. It requires many hours of considering many different avenues of attack, often requiring multiple perspectives from multiple experts to get the most accurate picture. In an extremely Agile-dominated industry where one of the key aims is to get a product out before competitors, there is simply not enough time in many cases to complete the STRIDE analysis to the highest possible level, limiting the quality of any output it produces.

Additionally, STRIDE generating so many novel ideas is both a blessing and a curse. It allows us to see many potential vectors for exploitation, though in the end many of these may end up being false positives. Also, depending on the quality of the threat modelling process itself, there may also be false negatives, which could be especially dangerous and lead to not only costs for the business, but also harm to consumer safety and privacy. Finally, STRIDE is very good at finding vectors for external threats, but is not so good at finding potential internal threats. All in all, it is a powerful tool, but only in the right contexts.

*3.2 CVSS*

As another popular threat modelling tool, it would also be beneficial to include CVSS in this research. We will use the most current edition at the time of writing, CVSS 3.1, as this will give the most comprehensive understanding of the current state of the tool and its applicability to this modern context.

A key detail about CVSS is that it assumes you already have a list of threats for which you want to calculate scores; because of this, we will be using the number of threats we have uncovered as part of our earlier STRIDE activities. Figure 8 below contains the results. Results have been colour-coded by level here for ease of overview. The numerical column on the left refers to the numbered threats from the STRIDE modelling. The Base Score gives the main calculation as given from FIRST's calculator, which is adjusted for contextual emergency to get a Temporal Score. This second score is adjusted to account for the physical circumstances (such as collateral damage potential with other devices) to gain the Environmental Score. The calculation of the latter two scores is generally considered optional in industry practice.

| No | Vector String | Base Score | Temporal Score | Environmental Score |
|---|---|---|---|---|
| 1 | CVSS:3.1/AV:P/AC:L/PR:N/UI:N /S:U/C:L/I:L/A:L/E:H/RL:U/RC: C/CR:L/IR:L/AR:L | 4.3 (M) | 4.3 (M) | 2.9 (L) |
| 2 | CVSS:3.1/AV:L/AC:H/PR:L/UI:N /S:U/C:H/I:H/A:N/E:U/RC:U/CR: H/IR:H | 6.3 (M) | 5.3 (M) | 5.9 (M) |
| 3 | CVSS:3.1/AV:L/AC:H/PR:L/UI:N | 6.4 (M) | 6.1 (M) | 7.4 (H) |
| | /S:C/C:N/I:H/A:L/E:P/RC:C/CR: M/IR:H/AR:M | | | |
| 4 | CVSS:3.1/AV:P/AC:L/PR:L/UI:R/ S:U/C:H/I:H/A:L/E:H/RL:U/RC:C /CR:H/IR:H/AR:L | 6.0 (M) | 6.0 (M) | 6.4 (M) |
| 5 | CVSS:3.1/AV:L/AC:H/PR:L/UI:R /S:U/C:N/I:H/A:N/E:U/RC:U/IR: M | 4.4 (M) | 3.7 (L) | 3.7 (L) |
| 6 | CVSS:3.1/AV:L/AC:H/PR:H/UI:R /S:C/C:N/I:H/A:L/E:U/RC:U/CR: M/IR:H/AR:L | 5.8 (M) | 4.9 (M) | 6.1 (M) |
| 7 | CVSS:3.1/AV:L/AC:H/PR:H/UI:R /S:C/C:L/I:H/A:L/E:F/RL:O/RC:C /CR:M/IR:H/AR:L | 6.4 (M) | 5.9 (M) | 6.8 (M) |
| 8 | CVSS:3.1/AV:L/AC:H/PR:L/UI:R /S:U/C:H/I:N/A:N/E:F/RL:U/RC: R/CR:H | 4.1 (M) | 4.4 (M) | 5.8 (M) |
| 9 | CVSS:3.1/AV:L/AC:L/PR:N/UI:R /S:U/C:L/I:N/A:N/E:U/RC:R/CR: M | 3.3 (L) | 2.9 (L) | 2.9 (L) |
| 10 | CVSS:3.1/AV:P/AC:L/PR:N/UI:R/ S:U/C:H/I:N/A:N/E:H/RL:U/RC: C/CR:H | 4.4 (M) | 4.3 (M) | 6.1 (M) |
| 11 | CVSS:3.1/AV:N/AC:L/PR:H/UI:N /S:U/C:H/I:N/A:N/E:H/RL:U/RC: C/CR:H | 4.9 (M) | 4.9 (M) | 6.7 (M) |
| 12 | CVSS:3.1/AV:P/AC:L/PR:N/UI:N /S:U/C:N/I:N/A:H/E:H/RL:U/RC: C/AR:M | 4.6 (M) | 4.6 (M) | 4.6 (M) |
| 13 | CVSS:3.1/AV:L/AC:H/PR:H/UI:N /S:U/C:L/I:N/A:H/E:P/RC:U/AR: M | 4.7 (M) | 4.1 (M) | 4.1 (M) |
| 14 | CVSS:3.1/AV:L/AC:H/PR:L/UI:N /S:U/C:N/I:N/A:H/E:U/RC:U/AR: M | 4.7 (M) | 4.0 (M) | 4.0 (M) |

| 15 | CVSS:3.1/AV:L/AC:H/PR:H/UI:N /S:C/C:H/I:H/A:H/E:P/RL:W/RC: C/CR:H/IR:H/AR:H | 7.5 (H) | 6.9 (M) | 7.0 (H) |
|----|------------------------------------------------------|---------|---------|---------|
| 16 | CVSS:3.1/AV:L/AC:H/PR:H/UI:N /S:C/C:H/I:H/A:H/E:P/RL:W/RC: C/CR:H/IR:H/AR:H | 7.5 (H) | 6.9 (M) | 7.0 (H) |

Figure 8: CVSS 3.1 table for the context of the home virtual assistant device

As before, several pros and cons become clear. Getting the scores is relatively simple as the numbers are computed by an official online calculator (FIRST, 2019). The understanding is also easily transferred; it is easy to explain which threats are potentially more severe than others because of the easy concept of finding the larger numbers. Additionally, the option to obtain three types of scores depending on how you are approaching the concept of what constitutes a 'threat' in the first place gives multiple perspectives on the same scenario.

As we can see, it is also easy to represent the CVSS data visually. In this case, colour coding makes it simple to pick out the highest risks from the mediums and lows. Perhaps the greatest strength of CVSS is this ability to ascribe scores in the first place. This allows stakeholders to prioritise, which as previously discussed is paramount in an industry so focused on speed of rolling out new products and updates.

On the other hand, CVSS only measures severity and not risk. In other words, while we can see which threats would have the most potentially-disastrous consequences were they exploited, we cannot tell what the actual likelihood of this happening may be. This might lead a manufacturer to focus efforts in the wrong place, trying to eliminate a threat which is almost impossible to create a reliable exploit for instead of the ones which are perhaps less devastating but more easily manipulated.

Another drawback is the fact that CVSS is subjective, perhaps surprisingly for a system based on numerical outputs. When designing the 'threat' to put through the calculator, one must make some subjective judgments about the way that threat is likely to take shape, which influences the final result. This means that the scores, while standardised in format, may differ depending on who is doing the activity. The solution of this would be to have many experts calculate the scores and then take averages, but this would multiply the time and effort required. This leads us to the final drawback - CVSS can only be used in conjunction with other threat modelling techniques, never alone. This means it can only add time to existing approaches, presenting an opportunity cost. Whether or not this is worthwhile depends on the priorities of the stakeholders.

### 3.3 Attack Trees

It would do us well to consider the oldest method of threat modelling, that being attack trees. In this approach, we are not required to use prior data from our other activities, and thus will not be doing so for the sake of fair comparison.

Three trees have been constructed, each with a top-level goal of compromising one of the CIA triad of information security. It should be noted that there are many conflicting opinions on which is the 'standard' or 'best' way to construct and label an attack tree; the format used here in Figures 9-11 is just one style of many.

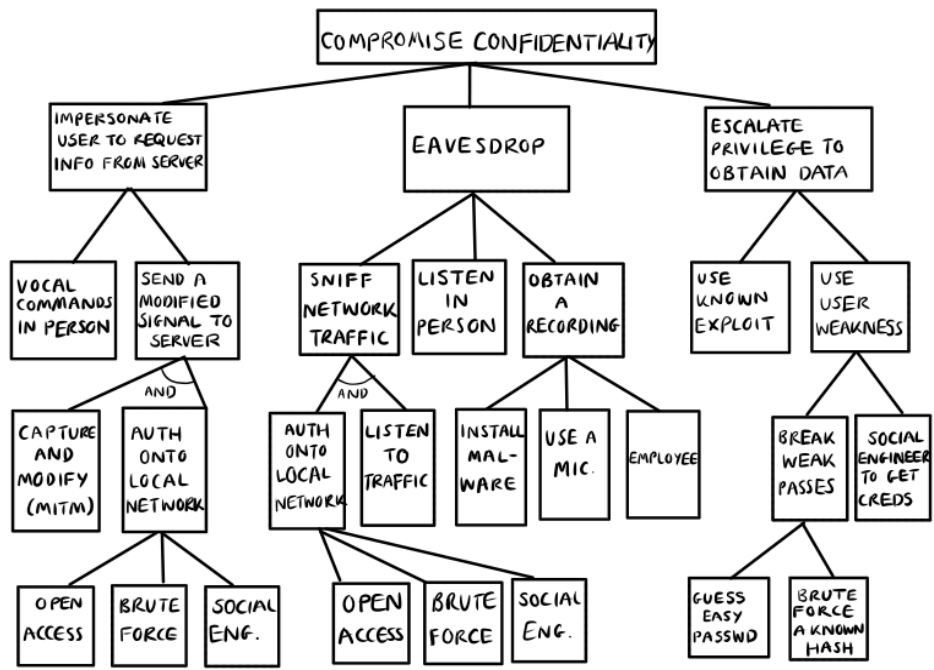

Figure 9: Threat tree of vectors for compromising confidentiality.

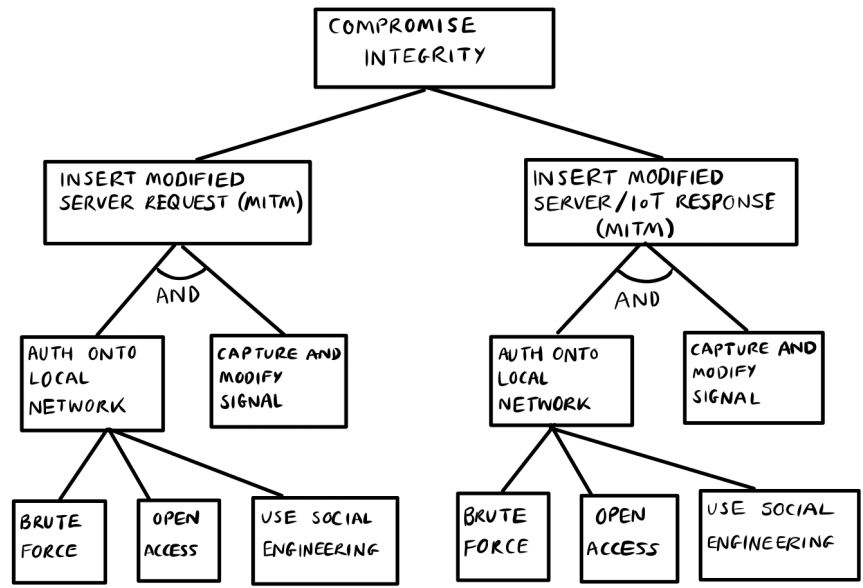

Figure 10: Threat tree of vectors for compromising integrity.

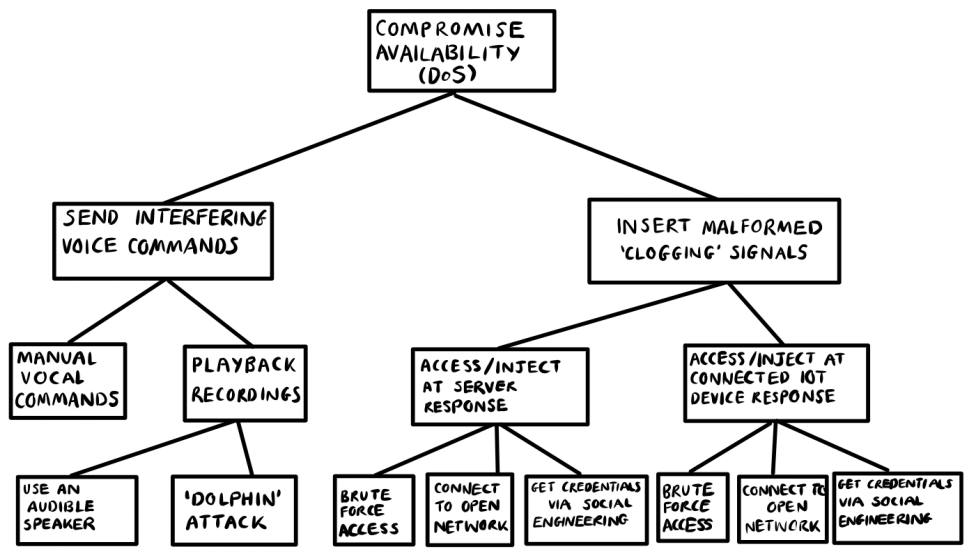

Figure 11: Threat tree of vectors for compromising availability.

One benefit of attack trees is that they can be as simple or complex as is desired. This means the time spent on the technique can be varied depending on how much time is available (advantageous in the case of the industry at hand) - though of course more thorough work almost always means more useful results. The many different ways of producing the trees also allows some flexibility on the part of the threat modeller. The outputs are well-presented visually, which also facilitates easy knowledge transfer. The final thing to note is that this technique helps us understand the particular methods attackers may use, thus allowing us to determine key access points or target processes where protective measures may best be focused. Taking all this into consideration, attack trees are quite an efficient approach in our scenario.

However, the variety of presentation methods can be a point of contention for some (how many nodes should be split, what level of detail is best, and so on). This can make it difficult to decide upon a best practice approach. As a result, it can be easy to over-complicate matters, thus diminishing the main advantages of attack trees being easily understood and efficient to make. Finally, in this particular device scenario, many of these nodes end up being somewhat repetitive as the steps to manipulate a threat can be quite similar, thus diminishing efficiency slightly, but not enough for this to be a significant problem. When considering all this, it may be most effective if the attack trees were combined with one other threat modelling approach in this context.

*3.4 LINDUNN GO*

After considering the oldest method in the threat modelling repertoire, we can now look at one of the newest. LINDUNN GO is a modern take on the activity, so from first glance we might expect it to be one of the most effective. Figure 12 shows the results of applying this framework to the home virtual assistant problem and then collating results in a table.

| GO Aspect | Details |
|---|---|
| Linkability of credentials | All user information is recorded to the identifying account created when installing the hub in the home. This may include credentials for other devices such as home security or smart energy systems. |
| Linkable user actions | -Specific queries or prior purchases can be collated in a user profile, which may manipulate advertisements, product pricing, and so on.<br>-Unauthenticated users interacting with the system in the home environment will also have their data collected even though they may not have signed any explicit consent agreements.<br>-Unauthenticated users can affect the aforementioned profiles of another user to whom the device belongs. |
| Linkability of inbound data | A small amount of transmitted personal information can be linked with pre-stored information in order to determine something larger about the user. |
| Linkabilty of context | Contextual data such as when the user turns an IoT home security system on or off can be used to glean a usage pattern, which may be used, for example, to determine when nobody is home in a particular house. |
| Linkability of shared data | Data collected is assumed to be added to a profile of the user. This increases the risk with the more data the user shares. May eventually lead to awareness or noncompliance problems. |
| Linkability of stored data | A company may only require the aggregated set of all data for analysis and service improvement purposes, but it may be the case that data is stored on a per-user basis, potentially making them highly identifiable to database intruders. Data should be aggregated across all users and minimized before any storage takes place. |
| Linkability of retrieved data | Related to above threats. Data may be retrieved that is identifiable to a specific user or users if correct minimization steps are not taken. A receiving party may also link more information together, the more is stored in this centralised data store. |
| Identifying credentials | Credentials in traffic and data storage allow identification of a user and their actions, which may be used for profiling. Is this always required (e.g. would Amazon really need to capture your IoT light switch usage and connect it to your credentials)? |
| Actions identify user | Repeated queries to a system can be used to compare (for gleaning identity) or profiling a user. Could it also be possible that someone visiting another home could have their location identified if their voice is stored on record with their own device? Other home users also contribute to the profile of the home user, which might lead to sabotage or incrimination. |
| Identifying inbound data | Unencrypted credentials could be used to identify a specific user. Those in the company who analyse voice recordings might be able to glean personal details from inbound data if it is not correctly cleaned before they gain access to it. |
| Identifying context | Even if a potential home invader does not know the specific identity of the person whose traffic they are sniffing, they may still be able to use data patterns (even encrypted!) to determine when a user is home. |
| Identifying shared data | If data is assumed to be transmitted anonymously for a certain purpose (e.g. service improvement) but quasi-identifiers may be collated to identify a person more clearly, this would be a privacy violation, potentially resulting in noncompliance. |
| Identifying stored data | See 'linkability of stored data'. |
| Identifying retrieved data | If returned data is insufficiently encrypted and contains identifying information, this can be sniffed. |
| Credentials non repudiation | Not applicable as this is only really a threat for systems for which usage is considered 'sensitive'. |
| Non repudiation of sending | Transmitted data can sometimes be accidental and highly sensitive (i.e. private conversations in the home). This may be linked to users or heard by analysts. |
| Non repudiation | Not applicable here. |
| of receipt | |
| Non reputable storage | Data in storage cannot easily be denied, which may be problematic if data is accidentally recorded (as is sometimes the case in this device context). However, this could be considered a good thing by some actors (such as if a crime is recorded). |
| Non repudiation of retrieved data | Data retrieved by a bad actor contains undeniable information, which could be used for blackmail or other purposes. |
| Detectable credentials | Not applicable as this is only really a threat for which having an account is considered 'sensitive'. |
| Detectable communication | See 'identifying context'. |
| Detectable outliers | A network intruder may be able to glean that a new device (potential theft target) has been added to the network by new and unusual traffic (even if encrypted, perhaps). Additionally, a new outlier might let them know that some other emergency is taking place that has been detected (e.g. break in or fire). |
| Detectable at storage | Not applicable here. |
| Detectable at retrieval | Not applicable here. |
| No transparency | Other people in the home are not sufficiently informed about the collection and use of their data. Users are also often ignorant about where their data goes as it is hidden behind lengthy walls of text or paperwork they don't read. |
| No user friendly privacy control | It can be hard for users to understand or change their privacy settings if they are not power users. Additionally, because of the above issue, many users aren't actually aware of things that they would otherwise like to change. It is generally no-privacy by default and the user has the burden of figuring out how and why to change this. |
| No access or portability | Not applicable here as it is a requirement. |
| No erasure | Not applicable here as it is a requirement. |
| Insufficient consent support | It can be very difficult to rescind consent to a certain data item in some cases. |
| Disproportionate collection | If data is collected about other devices in the connected home network, arguably this is not relevant and could be used for ulterior motives such as marketing which has not been specifically opted into. Additionally, it is questionable as to the motives where this is linked to identifying account data. |

| Unlawful processing | Users are generally required to opt in to data processing, but are generally unaware that this can involve people listening to their (accidental and intentional) recordings. This raises an ethical issue. Should users be made more specifically aware in agreements that other humans will often listen to their stored recordings? Additionally, it is hard to opt out of a newly introduced processing rule such as a new third party marketing allyship. |
|---|---|
| Disproportionate processing | Is personal data being used for training datasets (without consent) or employee training? Is it really necessary for employees to hear the recordings for data quality purposes? |
| Automated decision making | Not applicable here. |
| Disproportionate storage | Arguably (and depending on storage design), more data is stored than is necessary, such as connected devices and potential marketing topics. |

Figure 12: Results of application of LINDUNN GO to our context

One large advantage of using this approach becomes immediately obvious upon completing the activity. Privacy is a critically key concern in the context of home virtual assistant devices (and to an extent, IoT devices as a whole). As a result of this, the LINDUNN GO activity proves incredibly insightful and allows us to uncover potential issues which the other approaches have yet to discover.

The method of applying LINDUNN GO is also very straightforward due to it being broken down into manageable steps. It could be done by an individual, or by multiple people at once in a shorter amount of time. This gives valuable flexibility in a rapid-output industry. Additionally, the official documentation is arguably the most straightforward of any of the threat modelling methods covered thus far.

Something else to bear in mind is that the outputs of this activity are relatively accessible to the layperson who does not have a great deal of cyber security knowledge. This is one of the main things we are looking for, something that can be applied to educate device owners of the potential threats they may not be fully aware of. LINDUNN GO is the most powerful tool thus far in terms of how easily it can be accessed by or explained to the average consumer and grant them a greater degree of informed agency in their choice of home devices.

However, while this method is very good at uncovering privacy-related threats, it is not comprehensive at uncovering other types or telling us how particular exploits may be used by bad actors to compromise a system. Therefore, LINDUNN GO is best used in conjunction with one of the more traditional threat modelling techniques in this study, if we want to gain a clearer understanding of the whole picture when it comes to home digital assistants.

*3.5 Quantitative TMM*

Quantitative TMM is undertaken by combining the techniques of STRIDE, CVSS, and attack trees. Thus, for the sake of efficiency, figures will not be repeated in this text unnecessarily. Over the course of completing different threat modelling activities, the experience of undertaking Quantitative TMM has been gradually uncovered.

The approach takes the best of what these three threat modelling techniques have to offer; therefore we can easily say that Quantitative TMM is very powerful in the hands of sufficient expertise. The breadth of information that it uncovers is its primary strength. Furthermore, all of its components are well-known in industry, making it easy to transfer the knowledge among stakeholders, though this is perhaps limited by the exhaustive amount of information there is to get through. This is aided, however, by the way some aspects such as attack trees and CVSS are well-represented visually.

As we have just alluded to, though, Quantitative TMM is an extremely fine-toothed approach, and thus takes an extremely long time to carry out. STRIDE alone is a time-consuming approach, and adding two more threat modelling techniques on top of this makes for a high time and effort investment. This is totally at odds to the industry context we are working with; many top manufacturers do not have the time to undertake such activities when working to tight schedules. This severely limits how far we can recommend it as a best practice approach here.

Many aspects of the sub-activities within this approach are also subjective. The one-size-fits-all STRIDE technique also does not uncover many of the privacy implications that LINDUNN GO did, something which is critical in the IoT context. It would be better for us if it also incorporated some aspect of privacy threat analysis, though this would only compound the time-sink of the entire process.

*3.6 Findings*

Figure 13 gives us a matrix of each of the threat modelling activities we have applied, along with their key attributes.

| | Investment of time/effort | Prioritisation of issues | Scope | Visualisation / Accessibility | Purpose |
|---|---|---|---|---|---|
| **STRIDE** | High | Low | Varied | Complex matrix | Wide ranging threat discovery |
| **CVSS 3.1** | Med | High | Varied | Complex matrix | Severity analysis |
| **Attack Trees** | High | Low | Varied | Diagrams (varying) | Detailed, efficient attack vector discovery |
| **LINDUNN GO** | Med | Low | Focus | Simplified matrix | Privacy threat discovery |
| **Quantitative TMM** | Very High | High | Varied | Very complex, multi-part | Wide ranging threat discovery and analysis |

Figure 13: A summary of the applied threat modelling approaches

We have explored the best and worst aspects of each in terms of our context. We can make several statements about the context of home virtual assistant devices:

1. The industry deals with fast product development and update cycles, with an emphasis on maximum efficiency for time.

2. Threats are constantly evolving and increasingly sophisticated, across a wide variety of attack types.

3. Privacy threats are the most critical ones in the general public eye, as well as when one considers the amount of personal information processed by such devices.

As such, a combination of LINDUNN GO and STRIDE seems to provide the optimal chance to cover all threat-type bases. LINDUNN GO takes care of the integral privacy threat modelling and is done comparatively quickly, while STRIDE supplements it with the wider-reaching other categories of threats. This combination also gives outputs that can be partially understood by consumers, who arguably have a right to understand threats of devices they own or are considering purchasing. These approaches in tandem are less time-consuming as an entire Quantitative TMM undertaking, saving time in a time-focused industry. If the time opportunity or need for extra depth arises, though, CVSS could also reasonably be applied to our tandem approach relatively quickly in order to prioritise and categorise by severity, and the three would each complement the discoveries made by the others. This triad would provide a kind of equivalent to the Quantitative TMM approach which has the added advantage of exploring the privacy issues so intrinsic to our device type.

Therefore, this triad is what shall be used for the software development aspect of the study.

# 4. Requirements Analysis

## 4.1 Current Offerings

A brief overview of the primary threat modelling tools in the market is contained in Appendix A. From this, we can glean two main problems when considering the perspective of the home user. Firstly, the vast majority of these products are intended for corporate-level use, in networks with far more devices than the average smart home. As a result of this, output information is presented in business-centric terms which may not be wholly applicable, and additionally many of these products have a high price point which we can assume to be inaccessible or otherwise unjustifiable for many home users. The second drawback is that the above offerings are geared towards those who already have expertise in IT security, and are therefore rich with jargon and complex tools which would not suit the average smart home device user.

A gap in the current offerings can thus be gleaned: there is a need for a program which educates home users on the threats to their home IoT device network, in easy-to-understand terms, with no monetary cost so it is more accessible. In addition to this, any new software should also have the following qualities in order to be accessible to as many home users as possible:

- Compatibility with modern home releases of Windows and MacOS
- Small executable file size
- Processing power requirements which are as low as possible
- Partial sandboxing so that no data from the inputs is allowed to flow out of the software

## 4.2 Software Requirements and Use Cases

Figure 14 contains a basic context diagram for how this software may function.

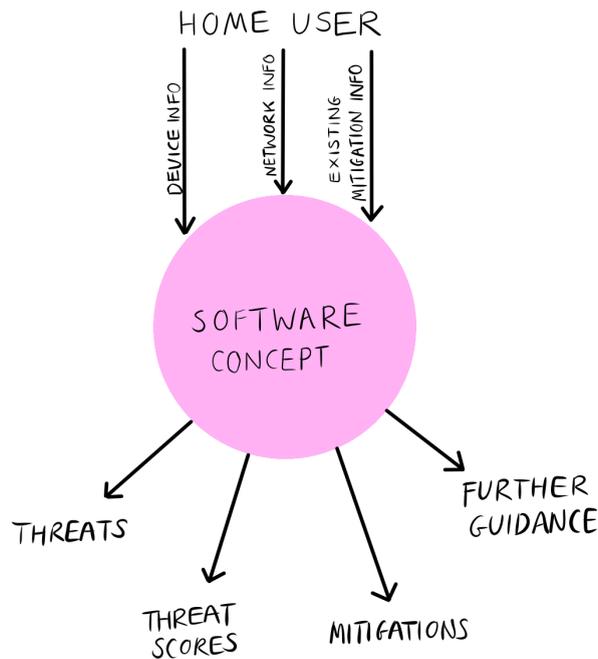

Figure 14: Context diagram for software concept.

With this concept in mind, a table was produced plotting what could feasibly be prototyped within the time frame of this project, based upon the MoSCoW method (Clegg & Barker, 1994).

| Must | <ul><li>Educate users on the potential risks to their home IoT network</li><li>Educate users on how to mitigate these aforementioned risks</li><li>Allow users to 'design' the state which closely matches their network, using checkboxes and questions</li><li>Use a graphical user interface</li><li>Be compatible with the majority of home tower/laptop computers (Windows 8 and up, equivalent MacOS release)</li></ul> |
|---|---|
| Should | <ul><li>Calculate an associated severity rating for each risk</li><li>Rank threats according to severity</li><li>Allow the user to account for any protective measures they already take in their initial design of the state of their network</li><li>Link users to where they can find further expert guidance based on the devices they own</li></ul> |
| Could | <ul><li>Have an attractive design which promotes user satisfaction and ease of use</li><li>Allow the user to personalise their experience by having the program address them by an input name</li><li>Have different colour schemes to combat eye strain and individual tastes</li></ul> |
| | • Dismiss threats in the list by clicking a button<br>• Allow users to print out or email the produced outputs |
| Would (but won't in this release) | • Allow the user to make a drag and drop visual network diagram akin to a simplified DFD<br>• Have a visual assistant (like Microsoft's Clippit) to promote ease of use<br>• Continually update to account for new devices and threats<br>• Take data from constantly updating threat databases to provide 'alerts' |

Figure 15: MoSCoW table for the software concept.

High-level requirements were gleaned as follows: to provide home IoT network threat modelling capabilities to end users, and to educate home IoT users on aspects of network and device security.

Figure 16 describes the typical assumed use case. A flow chart for this use case can be found in Figure 17 below.

| |
|---|
| *The user will be introduced with an introductory paragraph explaining how to use the software. They will then select from a list those IoT devices which are part of their home network. On the next page, they will indicate which devices directly interface with one another through checkboxes. Depending on the devices they selected, they may also be faced with some simple checkbox questions to determine what protective steps they are currently taking to secure their devices, if any. The software will then determine the key risks based on the user's inputs and assign each risk a severity rating. These risks will be presented to the user in a list, ranked by severity, and accompanied with what steps can be taken to mitigate said risk. Finally, the user will have the option to open a pop up menu which shows them where to find further guidance for the devices they own.* |

Figure 16: Use case for proposed software.

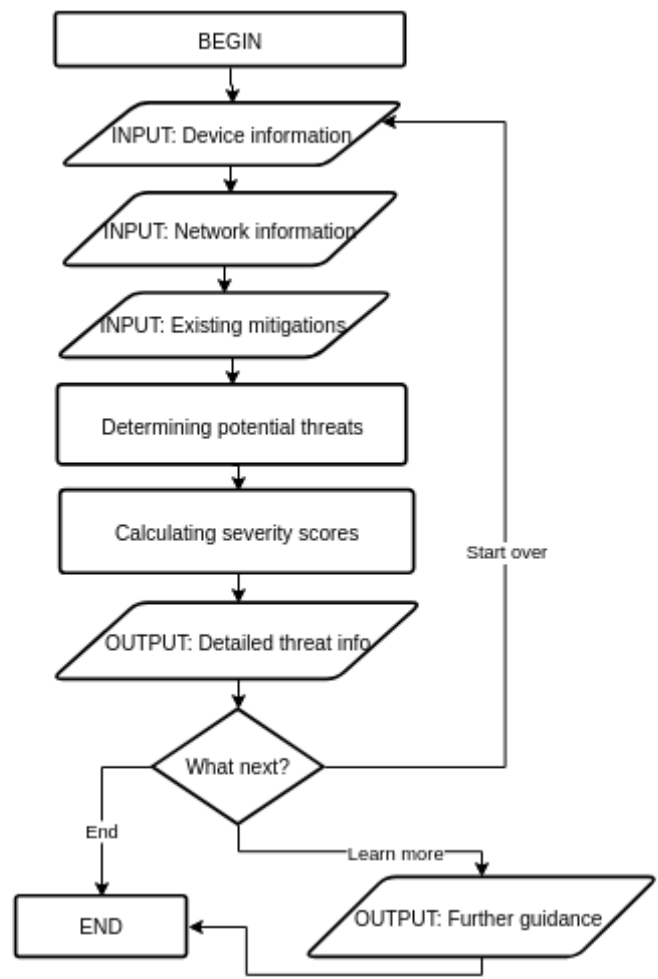

Figure 17: Use case illustrated as a flow chart

The following integration requirements were gleaned from this use case:

- For inputting network information: a Boolean value for whether each device in a potential list has been selected or not, internal names for each device, Boolean values for whether each mitigation or risk factor in a potential list has been selected or not, an internal name for each of these, checkboxes for toggling Boolean values, and a button which submits the selections.

- For calculating threats: an internal database of threat types, an internal database of scores assigned to various threat types, and an algorithm which creates permutations of the stored scores depending on the individual context information submitted by the user.

- For outputting information to the user: an output space which can be populated with the results of the algorithm, a button which links the user to further guidance, and a graphical user interface.

Figure 18 shows the elicited requirements as sorted into functional and non-functional categories.

| Functional | Non-functional |
|---|---|
| - User should be able to input devices owned<br>- User should be able to input device connections<br>- User should be able to input current protective techniques<br>- Software should identify risks based on inputs<br>- Software should calculate scores for each risk<br>- Software should output risks in order of severity score<br>- Software should link users to further guidance | - The software should be easy to understand for beginners<br>- The software should not transmit data about the user's inputs outside of the software itself<br>- The software should not significantly 'lag' or take up excessive processing power<br>- The software should be visually appealing<br>- Users should not be allowed to edit the internal calculations used to order threats<br>- Text should be easy to read |

Figure 18: Functional and non-functional requirements.

No requirement conflicts were found.

Rough concept diagrams were created at this stage; these are contained within Appendix B.

# 5. Project Management

*5.1 General Project Management*

Before undertaking any part of this project, a rough activity timeline was devised. The Gantt chart for this can be found below. To ensure appropriate progress and quality of outputs, regular meetings were scheduled with an academic adviser; the meeting log can be found in Appendix C. Thorough risk analysis and ethical approval stages were passed through before completing any work with volunteer participants.

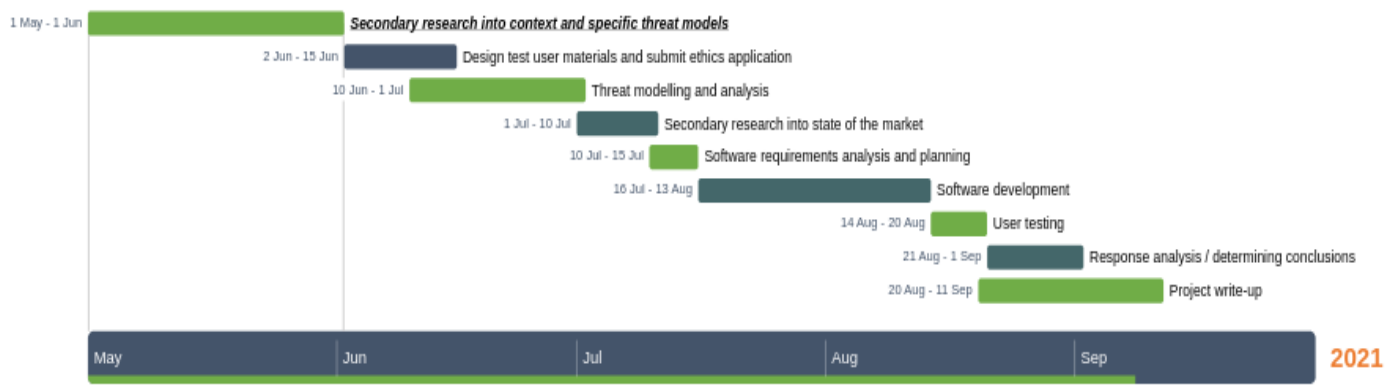

Figure 19: Gantt chart showing each project stage; note that there are some overlaps.

One discrepancy between the project and the proposal is in the amount of primary research participants. It was originally proposed to collect 10 test user responses, however, after researching the well-established literature on usability testing, 5 was reported to be the maximally efficient amount (Nielsen, 2012). This, combined with the increased difficulty in finding volunteers (and ensuring their safety along with the researcher's) during the current public health context, ultimately influenced the decision to go with 5 participants. An unforeseen positive of this is that it allowed more time to be devoted to polishing the software prototype, that would otherwise have been spent gathering and working with participants.

*5.2 Software Development*

An Agile-based approach with weekly sprints was selected for software development as this allowed for regular reflection on how to improve the quality of the work, thus hopefully creating a more successful overall output by the end of development. This approach was also selected as it allows for re-prioritisation or 'steering' of the project based on potential unforeseen circumstances.

Each weekly sprint was proposed to complete a new core feature or group of features from the specifications, with a stable software that would run smoothly and perform that particular task without error by the end of each sprint. After assessing the quality of the week's output, the next sprint would commence with any tweaks that were determined to be necessary, followed by the core feature(s) to be added next. A separate Gantt chart showing the tasks for each sprint can be found below.

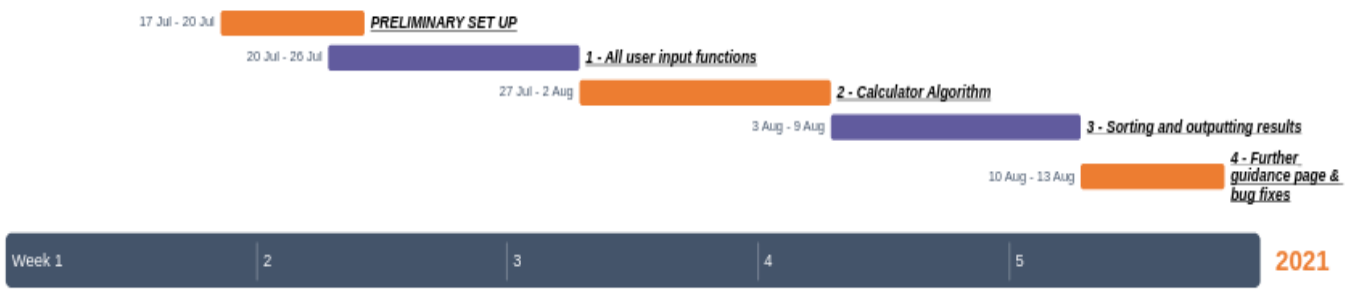

Figure 20: Gantt chart showing the tasks for each sprint

One drawback of this approach to development is that it became quite difficult to give subjective judgements about quality and required tweaks with only one developer, due to the limitations of having only one set of possible viewpoints. This was mitigated by using some of the regular project advisor meetings to discuss software progress, although this was not perhaps as accurate as having the opinions of other stakeholders, as the advisor was already an expert in the subject and therefore not the intended end user. However, this second opinion was better than having just one opinion, and ended up being very valuable in shaping the development outputs in the long term. Perhaps it may have been effective to seek ethical approval for further user testing stages, so that opinions could be tracked against progress towards the final prototype product. However, this would also have required a significant amount of time and paperwork, so the opportunity cost is somewhat limited in this respect. In any case, the processes that were used provided outputs that met the goals set out in requirements analysis, and performed well during user testing, so were arguably sufficient.

*5.3 Planning Stage*

The planning stages are covered at length in Section 4. As planning is arguably the most crucial of all stages in developing a strong software product, a great deal of time and care was put into this. The justification for this was that getting a strong grasp of the primary goals would make development more streamlined, having something to aim for with each sprint. Additionally, the investment of time in the beginning prevents the risk of having to spend more in the long term, if vague requirements and concepts have not been fleshed-out enough to produce high quality outputs.

It can be argued that this meticulous strategy was the most optimal one, as there was not a need to revise or add to requirements planning documentation after development had begun. Particularly, the combination of context diagrams, use cases, MoSCoW analysis, flow charts, concept designs, and sorting of requirements seems to have provided an effective spread in preparing to create a program and have metrics for its performance/efficacy.

*5.4 Design Stage*

One of the first key decisions was to take a minimalistic approach to visual design. User experience was deemed to be the most important gauge of success, and therefore clutter was considered something which could ultimately detract from that by way of distraction and confusion. Thus, the interface was planned to have a few simple buttons, minimal complex options, and a muted colour scheme. All of this was intended to promote a sense of 'relaxation' in the user, as the goal was to educate them about often-complex threats to their devices, something likely to be intrinsically stressful for many. The idea was, 'if the interface is relaxing and easygoing, then the concept of learning about device security will also be as easy as possible'.

Also part of this minimalist approach was the text contained within the program. A clean, sans serif font was chosen as these fonts are easiest to read and are not distracting. Sans serif fonts are also more accessible for users with dyslexia (Rello & Baeza-Yates, 2013). As for the text itself, the technical jargon was minimized and the GUI was presented in simple English. This is because the primary audience would not be experts, instead average consumers looking for a starting point for education. It became apparent during development that some very basic jargon (e.g. 'router') was unavoidable, so a simple glossary function was added to the software as a mitigation to improve ease of understanding.

The results of user testing show that this minimalist approach paid off (see Section 6) and the glossary in particular found use for testers, though there is still room for improvement in terms of balancing 'relaxing' and 'interesting', as we are to cover.

An idea from the concept designs included small product icons depending on what devices were selected by the user. These did not make it into the finished prototype for two reasons - firstly, the time required to source/create and implement all of these icons was deemed to be less important in the tight sprint schedule; secondly, they would have bloated the file size without adding much key functionality, something at odds with the desired specifications. Ultimately, this was arguably the correct decision as it allowed more time to be spent on more critical features, and product icons were not specifically requested as an additional feature by test users.

*5.5 Implementation Stage*

From the beginning it was decided that Python would be used, not only because of its simplistic and shorthand syntax, but also because it allows for the creation of elegant GUIs with the tkinter module. It is also powerful, allowing for fast creation of prototype builds. Furthermore, this is the language the developer had the most experience in, so was the most logical choice as opposed to grappling with lesser-known ones. A potential drawback of Python is its speed, but as a key specification was to keep the program as small and simple as possible, this was not an issue. Python turned out to be adequately powerful for all the program's needs.

The 'meat' of the coding came in the form of the scoring algorithm, which took elements from STRIDE, CVSS, and LINDUNN GO; these were deemed the most relevant for context from the literature review. CVSS in particular was integral to the rankings aspect. As the program was envisaged to give sufficiently accurate rankings, the scoring system ended up becoming very complex with many stages, all of which can be found in the algorithm documentation accompanying the program (a copy of this document is contained within Appendix D). Arguably, the algorithm was the most important component in the software, and a lot of time and revisions went into tweaking it for increased accuracy. For example, the very first pass at the algorithm did not lend as much weight to the supplied 'risk factors'; this was later changed to better reflect any mitigations users might already be taking with their devices. Ultimately, it is difficult to gauge how 'good' the algorithm is without a very high-level knowledge of all the associated threat modelling literature, so it is hard to say how much of a success this was. However, the gradual improvement of the system and positive user feedback from the results lend weight to the idea that it performed well.

One potential feature which did not make it into the final prototype was the option for users to select a Bitdefender BOX as one of their devices; this is a device marketed as a 'full cyber security ecosystem for [the] home', which users connect to their home network with the aim of better protecting their other devices (Bitdefender, 2021). The primary reason for not including this at the final prototype stage was the fact that it would have functioned in the opposite way to the other devices on the list, by removing risks and subtracting from scores instead of adding to them. After frequent revisions to the already-complex algorithm, it was deemed that this inclusion would be difficult to achieve under project time constraints. Furthermore, it was considered unlikely that the intended user group would use one of these devices, as they tend to be used by people with more interest and knowledge in device security. This turned out to be a reasonable judgement as none of the testers had a BOX, and its lack of inclusion as an option was not missed.

# 6. Testing

## 6.1 Methodology

The software was then trialled by test users in order to determine its functionality, ease of use, and any unexpected occurrences; in essence, testing it against the criteria for success that were set out in the design stages.

Subjects were chosen based on a number of criteria. First of all, as the project focuses on UK consumers of digital assistant hub devices, the participants needed to be based in the UK, aged 18 or over, and based within a household with one of these devices. Additionally, due to the nature of testing, participants needed to be able to use a visual GUI-based program designed for those with a reasonable grasp of English. Similarly, they would also need to be able to reasonably communicate their thoughts afterwards, in English, to a researcher. Due to the public health situation surrounding COVID-19, access to participants was somewhat limited. As a result of this, testing recruitment was limited to academic colleagues, family friends, and friends of friends. Close friends were avoided due to the ethical implications of potential feelings of duress or obligation to give positive feedback; this was done with the aim of keeping results more accurate and credible. Based upon diminishing returns on efficiency, the optimal number of participants in usability testing is considered to be around 5; therefore, this was the number of participants chosen (Nielsen, 2012).

Each user test contained two parts. Firstly, the subject was given 10-15 minutes to fully explore the software, and encouraged to follow the 'think-aloud' protocol. The accompanying researcher made notes of any verbal feedback given at this point, any prolonged pauses or confusion on the part of the user, as well as any errors or unexpected occurrences arising with the software itself. This section was not video recorded due to the potentially sensitive nature of users inputting information regarding device security. The second part of the test involved a recorded interview where the user answered some open-ended questions about their experience, and then answered some scaled statements based upon Nielsen's usability heuristics (Nielsen, 2019). Some demographic (but non-identifying) data concerning age range and gender was also collected for each participant, so as to compare the tester population against the general population to ascertain generalisability. This was considered important, as a common limiting factor of user testing efficacy is skewed sampling.

A naturalistic observation style was used to increase the likelihood of the user acting as they 'truly would' in a home scenario with such a software. This style of observation also overcomes gaps between how a user believes they may act (and how they relay this information to a researcher), and how they actually act. The act of following up immediately with questions also allows for more accurate user recall, and therefore more detailed and accurate answers. Potential drawbacks for this method include a lack of 'natural' outside influences, and having only a single pass at data collection. In this scenario, users would be testing software designed for use in the home, so unexpected outside influences would likely be minimal. Also, a structured plan was designed to increase the likelihood of testing going smoothly the first time.

A full ethical risk assessment for the user testing phase is contained in Appendix F.

*6.2 Demographics*

Figure 21 is informed by research into the age demographics of home IoT devices and gives us an idea on the general makeup of the target population (Sanderson, 2018). A key difficulty in user testing is the small group size required for maximum efficiency, which therefore makes it difficult to accurately represent the demographics of a complex group. To mitigate this as much as possible, participants from either end of the age spectrum were involved. The data may have ended up being slightly more representative had there been a participant from either the 35-44 or 45-54 age groups, though the constraints on access to participants made this impossible in the time frame of this study. Ultimately, this test group is reasonably representative for the purposes of this activity.

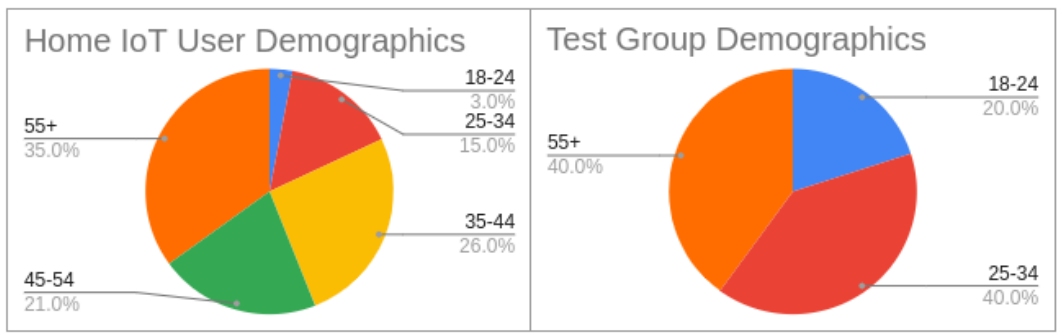

Figure 21: Comparing the age demographics of the general target population with the testing population.

Figure 22 is informed by statistics regarding gender demographics in the UK. As data regarding the gender demographics for home IoT users was unavailable, the general population was taken instead as the point of reference (ONS, 2011). The data used has a couple of considerable drawbacks, however. Firstly, the most recent census data available at the time of writing dates back to 2011, limiting its potential accuracy to a 2021 context. Furthermore, the only options for gender on the census were 'male' and 'female', which does not accurately represent the diversity of options as identified by test group participants. Therefore, the usefulness of the census dataset is limited. Again, the small group limited how well the test users could accurately represent the entire makeup of the target population, although the lean towards a more female population was reflected, though exaggerated somewhat (51% up to 60%) by sample size.

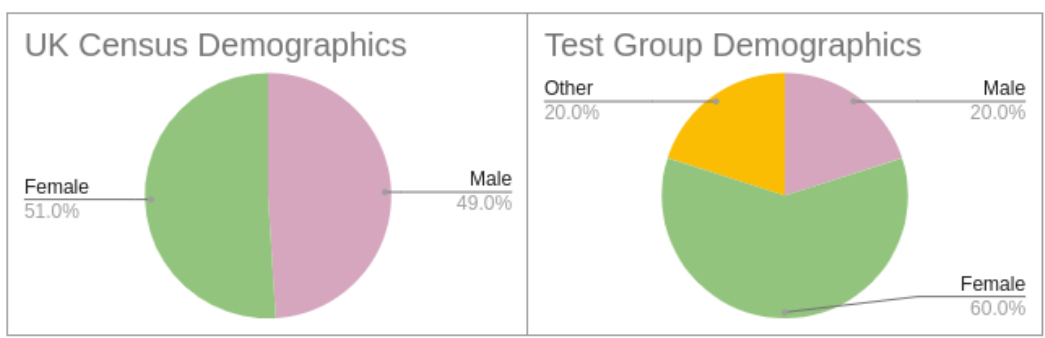

Figure 22: Comparing gender demographics of the general target population with the testing population.

Taking the limitations of the population data into consideration alongside the small sample size required, we can assume that this test user group provides an acceptable insight in terms of how generalisable it is, though naturally we cannot be wholly confident in the strength of statistical claims.

*6.3 Results*

None of the test participants experienced any significant pauses or noticeable confusion during the exploratory phase, suggesting that the software is straightforward to use. Furthermore, there were no instances of errors or unexpected program behaviours; this shows that there are likely no significant 'bugs' or problems with how it functions.

Most questions were asked by the participants from the '55+' age bracket, and related more so to IoT devices in general than anything within the program: 'what is the difference between a smart security camera and a smart doorbell' and 'what is smart kitchenware' are notable examples. These do not suggest any glaring problems with the software itself, though it perhaps highlights a need to put these definitions into the included glossary tool. One user asked a question which was answered by the accompanying user guide documentation, and they were satisfied with the answer they found, suggesting that the documentation is sufficient for this purpose.

Participants generally only used the 'think-aloud' protocol to voice minor opinions, such as regarding the colour scheme of the program or how informative it was. The vast majority of these were positive statements, with only one being less so, with regards to the layout, but this was further expanded upon in the interview segment which we will cover below.

Regarding the visual 'feel' of the software, users were generally positive. One remarked that the design could have been more engaging or interesting to look at, suggesting that this could be worked on in future updates to make the user experience more enjoyable. However, the same user also remarked that the software was 'fine for a prototype stage', suggesting that it is appropriate for its current form.

When discussing what they liked the most, 60% remarked on the 'thoroughness' or how 'specific' the outputs were. This points to the program fulfilling its primary goal of informing users about specific risks to their devices without being overly-confusing or full of obtuse jargon. 40% remarked that the software was 'easy' or 'friendly' to use, further backing up the claim that it is sufficiently oriented towards the non-expert demographic. One participant specifically pointed out the inclusion of severity rankings as something they liked.

In terms of what they liked the least, participants on the whole had less to say, though this could have been due to feelings of politeness more so than an actual lack of ideas. The participant that had remarked about the interface not being engaging brought this up in more detail here. One other stated that the inclusion of arrow buttons instead of using the mouse to scroll would have enhanced their user experience; this is a valid idea that would be relatively easy to implement in a future version and should be strongly considered.

As for other features participants thought would enhance the software, ideas included a helping mascot in the same vein as Microsoft's Clippit (something which was considered during planning stages but relegated to a 'future' idea due to time constraints), feedback sounds for button presses, and the ability to save a copy of the threat model outputs to return to later (something which was also considered a future idea in the planning stages). These responses suggest that a saving or export feature and a mascot would be potentially well-received future expansions and should therefore be considered. The mascot in particular may combat the feelings among certain participants that the user interface could be more engaging. Feedback sounds would also be a potentially good idea as they would make the software more accessible and interactive, theoretically enhancing the user experience. One participant also remarked that the program should be regularly updated to account for expanding knowledge of device types and threats; this is something that would arguably be done anyway in a full-scale version of the project but is nonetheless important to reiterate.

100% remarked that they would consider using a fully-developed version of the software if freely available, with 40% stating that the lack of a monetary cost influenced this decision. This is critical as one of the key aims was to create a program that would be usable for free to fill a gap in the market, and results suggest the response to such an offering would be positive. All of the respondents also remarked that they would recommend the software to other acquaintances, further backing up this idea.

Responses to the heuristics-based questions are included in Figure 23 below. As previously discussed, these statements were rated on a scale and were loosely based on Nielsen's usability heuristics, though slightly modified to be more applicable to this specific prototyping context.

| Statement | Strongly agree | Agree | Unsure / Neutral | Disagree | Strongly disagree |
|---|---|---|---|---|---|
| The software gave quick feedback to my inputs. | 100% (5) | 0 | 0 | 0 | 0 |
| It was obvious what I needed to do next at all times. | 100% (5) | 0 | 0 | 0 | 0 |
| I can see how I would be able to do it again if I wanted to. | 100% (5) | 0 | 0 | 0 | 0 |
| The software was consistent in what it did, and not unpredictable. | 60% (3) | 40% (2) | 0 | 0 | 0 |
| Information provided was of appropriate detail and length. | 60% (3) | 40% (2) | 0 | 0 | 0 |
| I found the software simple to use. | 100% (5) | 0 | 0 | 0 | 0 |
| I could understand what the software was telling me. | 80% (4) | 20% (1) | 0 | 0 | 0 |
| After using the software, my understanding of security threats to a home network has increased. | 80% (4) | 20% (1) | 0 | 0 | 0 |
| After using the software, I know some steps one might take to better secure their electronic devices. | 80% (4) | 20% (1) | 0 | 0 | 0 |
| Using this software, I could learn more about device security in general. | 80% (4) | 20% (1) | 0 | 0 | 0 |

Figure 23: Responses to heuristics-based questions.

These responses indicate strongly positive results. The areas where performance was best included responsive status, ease of understanding, ease of use, and the ability to recall what to do again. These are strong evidence towards the software fulfilling its goal of being simple for non-experts to use, and together all of these results suggest a strongly positive user experience.

*6.4 Potential Limitations*

As previously discussed, sample size is a limiting factor in the generalizability of results, though effort has been taken to ensure that participants from a wide range of demographic groups were included, decreasing the risk of bias. Something worth considering is that directly interviewing participants can sometimes lead to exaggeration of positive responses out of a feeling of politeness, though this is a risk intrinsic to any study of this type and must be accepted. Overall, the data collected is useful in determining what the next steps of development should be and in determining how well the development objectives are currently being met.

However, these results should not be taken as the entire breadth of possible user opinions and ideas.

# 7. Conclusions

From feedback analysis, we can gauge that the resulting product is usable, despite it still being a proof-of-concept. We have, however, received feedback that can be used to improve the usability and user experience. These should be actioned in future versions, or taken on board by anyone seeking to make a similar software.

The methods employed throughout development worked well considering the scope and time scale of the project. One of the key factors that likely contributed to success was the opportunity to meet regularly with the project supervisor, as this promoted regular progress reviews and target setting. While the primary data collection methodology was imperfect, it was arguably the most efficient and feasible approach given project restraints. If more time for the project was allotted, a more polished and complete software could have been produced, and testing could have been expanded to have multiple stages (perhaps on a per-development-milestone basis) and more participants to account for the more sophisticated system.

Potential alternatives to achieving the goals set out by the software include things such as training initiatives and awareness campaigns. However, these require users to set aside more significant amounts of time to devote to learning, whereas a desktop application is easily accessible while multitasking, and the prototype built does not take long to navigate. This combined with the positive user feedback means that we can reasonably consider this software education strategy optimal.

The software fulfilled all of the criteria in the 'Must' and 'Should' sections of the MoSCoW analysis. It was also relatively small, had low computational intensity, and did not require data flows out of the trust boundary of the program itself. Therefore we can argue that the core requirements were all met.

The Agile-based approach proved extremely valuable as a way of encouraging regular progress; the meetings with a project supervisor were particularly useful for assessing the strengths and points for improvement at each stage. The only notable drawback was that certain development stages took slightly more or less than a week to complete, and a rigid Agile schedule makes it difficult to account for this in planning and progress documentation.

The threat modelling process undertaken prior to software development found solutions which arguably worked best given the specific scenario. If the environment surrounding virtual assistant hub devices changes significantly, the optimal threat modelling process is also likely to change. For example, if the rate of new product release slows, then time will no longer be such a heavy factor in choosing methodologies, and the optimal strategy may change.

In terms of the industry value of these conclusions, they may be used to allow product developers to find an optimal threat modelling strategy that maximises depth and efficiency given a limited time frame. They may also be used to gain insight into where their consumers are lacking in knowledge, so that they know where they may improve communications to remedy this. Furthermore, any stakeholder interested in increasing information security awareness among the general population may take the software prototype and modify or add to it freely. Alternatively, they may use it as inspiration for their own projects made from scratch.

The ethical implications of this work are also significant. Arguably, some interesting insights have been gained into consumer perceptions of devices and trust levels. Additionally, the concept of promoting better cyber security in the home is founded in individual rights to privacy and security; with better knowledge, consumers and their data can become safer, lowering rates of cyber crime and accidental breaches.

Taking the industry value and current ethical implications of the research deliverables together, the project was, overall, rewarding and enlightening.

# 8. References

- OWASP, 2021. *Threat Modeling | OWASP*. [online] Owasp.org. Available at: <https://owasp.org/www-community/Threat_Modeling> [Accessed 13 May 2021].
- Oracle, 2021. *What is the Internet of Things (IoT)?*. [online] Oracle.com. Available at: <https://www.oracle.com/internet-of-things/what-is-iot/> [Accessed 2 September 2021].
- Hern, A., 2019. Apple contractors 'regularly hear confidential details' on Siri recordings. *The Guardian*, [online] Available at: <https://www.theguardian.com/technology/2019/jul/26/apple-contractors-regularly-hear-confidential-details-on-siri-recordings> [Accessed 3 May 2021].
- Shevchenko, N., Chick, T., O'Riordan, P., Scanlon, T. and Woody, C., 2018. *Threat Modeling: A Summary of Available Methods*. [online] Pittsburgh: Carnegie Mellon University, pp.1-19. Available at: <https://resources.sei.cmu.edu/library/asset-view.cfm?assetid=524448> [Accessed 1 May 2021].
- *Evaluation of Threat Modeling Methodologies*. 2016. [video] Directed by F. Shull. Carnegie Mellon University.
- Abomhara, M., Køien, G. and Gerdes, M., 2015. A STRIDE-Based Threat Model for Telehealth Systems. In: *Norsk informasjonssikkerhetskonferanse (NISK2015)*. Ålesund: ResearchGate, pp.1-15.
- FIRST, 2019. *CVSS v3.1 Specification Document*. [online] FIRST — Forum of Incident Response and Security Teams. Available at: <https://www.first.org/cvss/specification-document> [Accessed 3 September 2021].
- Scarfone, K. and Mell, P., 2009. An analysis of CVSS version 2 vulnerability scoring. *2009 3rd International Symposium on Empirical Software Engineering and Measurement*,.
- Amoroso, E., 1994. *Fundamentals of computer security technology*. 1st ed. Englewood Cliffs, N.J.: Prentice-Hall International, Inc.
- Salter, C., Saydjari, O., Schneier, B. and Wallner, J., 1998. *Toward a Secure System Engineering Methodology*. New Security Paradigms Workshop. [online] Schneier on Security, pp.2-10. Available at: <https://www.schneier.com/academic/archives/1998/09/toward_a_secure_syst.html> [Accessed 1 May 2021].
- Aufner, P., 2019. The IoT security gap: a look down into the valley between threat models and their implementation. *International Journal of Information Security*, 19(1), pp.3-14.
- DistriNet Research Group, 2021. *LINDUNN GO*. [online] Linddun.org. Available at: <https://www.linddun.org/go> [Accessed 3 September 2021].
- Nurse, J., Creese, S. and De Roure, D., 2017. *Security risk assessment in Internet of Things systems*. [online] Oxford: University of Oxford, p.7. Available at: <https://www.cs.ox.ac.uk/files/9680/2017-itpro-ncd_author-final.pdf> [Accessed 3 September 2021].

- Rizvi, S., Pipetti, R., McIntyre, N., Todd, J. and Williams, I., 2020. Threat model for securing internet of things (IoT) network at device-level. *Internet of Things*, 11, p.1-21.

- Pacheco, J. and Hariri, S., 2016. IoT Security Framework for Smart Cyber Infrastructures. In: *1st International Workshops on Foundations and Applications of Self\* Systems*. [online] Augsberg: IEEE, pp.242-247. Available at: <https://ieeexplore.ieee.org/abstract/document/7789475> [Accessed 1 May 2021].

- Bugeja, J., Jacobsson, A. and Davidsson, P., 2017. An analysis of malicious threat agents for the smart connected home. In: *IEEE Annual Conference on Pervasive Computing and Communications Workshops (PerCom)*. IEEE, pp.557-562.

- Cho, G., Choi, J., Kim, H., Hyun, S. and Ryoo, J., 2018. Threat Modeling and Analysis of Voice Assistant Applications. In: *International Workshop on Information Security Applications*. [online] Jeju Island: Springer, pp.197-209. Available at: <https://link.springer.com/chapter/10.1007/978-3-030-17982-3_16> [Accessed 1 May 2021].

- Lit, Y., Kim, S. and Sy, E., 2021. A Survey on Amazon Alexa Attack Surfaces. In: *3rd International Workshop on Security Trust and Privacy for Emerging Cyber Physical Systems*. [online] IEEE, pp.1-7. Available at: <https://ieeexplore.ieee.org/document/9369553> [Accessed 12 May 2021].

- Seeam, A., Ogbeh, O., Guness, S. and Bellekens, X., 2019. Threat Modeling and Security Issues for the Internet of Things. In: *International Conference on Next Generation Computing Applications (NextComp)*. Maritius: IEEE, pp.1-8.

- Hu, H., Yang, L., Lin, S. and Wang, G., 2020. A Case Study of the Security Vetting Process of Smart-home Assistant Applications. In: *2020 IEEE Security and Privacy Workshops (SPW)*. [online] San Francisco: IEEE, pp.76-81. Available at:
- <https://ieeexplore.ieee.org/abstract/document/9283882> [Accessed 1 May 2021].

- Jackson, R. and Camp, T., 2018. Amazon Echo Security: Machine Learning to Classify Encrypted Traffic. In: *27th International Conference on Computer Communication and Networks (ICCCN)*. [online] IEEE, pp.1-10. Available at: <https://ieeexplore.ieee.org/abstract/document/8487332> [Accessed 13 May 2021].

- Lei, X., Tu, G., Liu, A., Li, C. and Xie, T., 2018. *The Insecurity of Home Digital Voice Assistants – Vulnerabilities, Attacks and Countermeasures*. [online] Michigan State University, pp.1-9. Available at: <https://www.cse.msu.edu/~xietian1/paper/HDVA_Security-v2.pdf> [Accessed 13 May 2021].

- Haack, W., Severance, M., Wallace, M. and Wohlwend, J., 2017. *Security Analysis of the Amazon Echo*. [online] MIT, pp.1-14. Available at: <https://courses.csail.mit.edu/6.857/2017/project/8.pdf> [Accessed 13 May 2021].

- Alhabi, R. and Aspinall, D., 2018. *An IoT Analysis Framework: An Investigation of IoT Smart Cameras' Vulnerabilities*. Edinburgh: University of Edinburgh, pp.1-10.

- Zeng, E., Mare, S. and Roesner, F., 2017. End User Security and Privacy Concerns with Smart Homes. In: *Thirteenth Symposium on Usable Privacy and Security (SOUPS 2017)*. [online] Santa Clara: Usenix, pp.65-80. Available at:

<https://www.usenix.org/system/files/conference/soups2017/soups2017-zeng.pdf> [Accessed 1 May 2021].

- Kavallieratos, G., Gkioulos, V. and Katsikas, S., 2019. Threat analysis in dynamic environments: The case of the smart home. In: *15th International Conference on Distributed Computing in Sensor Systems*. [online] Santorini: IEEE, pp.234-240. Available at: <https://ieeexplore.ieee.org/document/8804734> [Accessed 12 May 2021].

- Zeng, E., Mare, S. and Roesner, F., 2017. End User Security and Privacy Concerns with Smart Homes. In: *Thirteenth Symposium on Usable Privacy and Security (SOUPS 2017)*. [online] Santa Clara: Usenix, pp.65-80. Available at: <https://www.usenix.org/system/files/conference/soups2017/soups2017-zeng.pdf> [Accessed 1 May 2021].

- Blythe, J., Sombatruang, N. and Johnson, S., 2019. What security features and crime prevention advice is communicated in consumer IoT device manuals and support pages?. *Journal of Cybersecurity*, [online] 5(1), pp.1-10. Available at: <https://academic.oup.com/cybersecurity/article/5/1/tyz005/5519411?searchresult=1> [Accessed 12 May 2021].

- O'Donnell, L., 2019. *Ring Doorbell Flaw Opens Door to Spying*. [online] Threatpost.com. Available at: <https://threatpost.com/ring-doorbell-flaw-opens-door-to-spying/142265/> [Accessed 3 May 2021].

- Jackson, R. and Camp, T., 2018. Amazon Echo Security: Machine Learning to Classify Encrypted Traffic. In: *27th International Conference on Computer Communication and Networks (ICCCN)*. [online] IEEE, pp.1-10. Available at: <https://ieeexplore.ieee.org/abstract/document/8487332> [Accessed 13 May 2021].

- Haack, W., Severance, M., Wallace, M. and Wohlwend, J., 2017. *Security Analysis of the Amazon Echo*. [online] MIT, pp.1-14. Available at: <https://courses.csail.mit.edu/6.857/2017/project/8.pdf> [Accessed 13 May 2021].

- Chung, H., Iorga, M., Voas, J. and Lee, S., 2017. "Alexa, Can I Trust You?." *IEEE*, [online] (9), pp.100-104. Available at: <https://www.computer.org/csdl/magazine/co/2017/09/mco2017090100/13rRUyoyhHl> [Accessed 12 May 2021].

- Griffin, A., 2018. Amazon Echo sends long recording of couple's private conversation to random person. *The Independent*, [online] Available at: <https://www.independent.co.uk/life-style/gadgets-and-tech/news/amazon-echo-alexa-record-conversation-privacy-security-a8368231.html> [Accessed 3 May 2021].

- Verheyden, T., Baert, D., Van Hee, L. and Van Den Heuvel, R., 2019. *Google employees are eavesdropping, even in your living room, VRT NWS has discovered*. [online] VRT NWS. Available at: <https://www.vrt.be/vrtnws/en/2019/07/10/google-employees-are-eavesdropping-even-in-flemish-living-rooms/> [Accessed 3 May 2021].

- Flikkema, P. and Cambou, B., 2017. When things are sensors for cloud AI: Protecting privacy through data collection transparency in the age of digital assistants. In: *Global*

*Internet of Things Summit (GIoTS)*. [online] IEEE, pp.1-4. Available at: <https://ieeexplore.ieee.org/document/8016284> [Accessed 12 May 2021].

- Hern, A., 2019. Apple contractors 'regularly hear confidential details' on Siri recordings. *The Guardian*, [online] Available at: <https://www.theguardian.com/technology/2019/jul/26/apple-contractors-regularly-hear-confidential-details-on-siri-recordings> [Accessed 3 May 2021].

- Clegg, D. and Barker, R., 1994. *Case Method Fast-Track: A RAD Approach*. Wokingham (UK) [etc.]: Addison-Wesley.

- Nielsen, J., 2012. *How Many Test Users in a Usability Study?*. [online] NN Group. Available at: <https://www.nngroup.com/articles/how-many-test-users/> [Accessed 24 August 2021].

- Rello, L. and Baeza-Yates, R., 2013. *Good Fonts for Dyslexia*. [online] Washington, p.1. Available at: <http://dyslexiahelp.umich.edu/sites/default/files/good_fonts_for_dyslexia_study.pdf> [Accessed 27 August 2021].

- Bitdefender, 2021. *Box Homepage*. [online] Bitdefender.co.uk. Available at: <https://www.bitdefender.co.uk/box/> [Accessed 27 August 2021].

- Nielsen, J., 2019. *Heuristic Evaluation of User Interfaces*. [online] NN Group. Available at: <https://www.nngroup.com/videos/heuristic-evaluation/> [Accessed 24 August 2021].

- Sanderson, T., 2018. *The Rise of Smart Tech in UK Homes*. [online] Thomas Sanderson. Available at: <https://www.thomas-sanderson.co.uk/inspiration/the-rise-of-smart-tech-in-uk-homes/> [Accessed 25 August 2021].

- ONS, 2011. *2011 Census*. [online] Office for National Statistics. Available at: <https://www.ons.gov.uk/census/2011census> [Accessed 25 August 2021].

# 9. Bibliography

- OWASP, 2021. *Threat Modeling - OWASP Cheat Sheet Series*. [online] Cheatsheetseries.owasp.org. Available at: <https://cheatsheetseries.owasp.org/cheatsheets/Threat_Modeling_Cheat_Sheet.html> [Accessed 2 September 2021].
- NJCCIC, 2017. *BrickerBot - NJCCIC Threat Profile*. [online] Cyber.nj.gov. Available at: <https://www.cyber.nj.gov/threat-center/threat-profiles/botnet-variants/brickerbot> [Accessed 3 May 2021].
- Herzberg, B., Zeifman, I. and Bekerman, D., 2016. *Breaking Down Mirai: An IoT DDoS Botnet Analysis*. [online] Imperva. Available at: <https://www.imperva.com/blog/malware-analysis-mirai-ddos-botnet/?redirect=Incapsula> [Accessed 3 May 2021].
- Kang, E., 2014. *Threat Modeling Made Interactive - OWASP AppSecUSA 2014*.
- Amoroso, E., 1994. Fundamentals of computer security technology. 1st ed. Englewood Cliffs, N.J.: Prentice-Hall International, Inc.
- Alberts, C., Dorofee, A., Stevens, J. and Woody, C., 2003. *Introduction to the OCTAVE® Approach*. Networked Systems Survivability Program. [online] Pittsburgh: Carnegie Mellon University, pp.1-27. Available at: <http://itgovernanceusa.com/files/Octave.pdf> [Accessed 1 May 2021].
- *Eddington, M., Larcom, B. and Saitta, E., 2005. Papers. [online] octotrike.org. Available at: <https://www.octotrike.org/papers> [Accessed 1 May 2021].*
- *Johnson, P., Lagerström, R. and Ekstedt, M., 2018. A Meta Language for Threat Modeling and Attack Simulations. In: ARES 2018: International Conference on Availability, Reliability and Security. [online] Hamburg: ACM, pp.1-8. Available at: <https://dl.acm.org/doi/abs/10.1145/3230833.3232799> [Accessed 1 May 2021].*
- *Morana, M., 2012. PASTA Process for Attack Simulation and threat analysis (PASTA) Risk-centric Threat Modeling. [online] Secure Software. Available at: <https://securesoftware.blogspot.com/2012/09/rebooting-software-security.html> [Accessed 1 May 2021].*
- O'Donnell, L., 2019. *Ring Doorbell Flaw Opens Door to Spying*. [online] Threatpost.com. Available at: <https://threatpost.com/ring-doorbell-flaw-opens-door-to-spying/142265/> [Accessed 3 May 2021].
- Herzberg, B., Zeifman, I. and Bekerman, D., 2016. *Breaking Down Mirai: An IoT DDoS Botnet Analysis*. [online] Imperva. Available at: <https://www.imperva.com/blog/malware-analysis-mirai-ddos-botnet/?redirect=Incapsula> [Accessed 3 May 2021].
- NJCCIC, 2017. *BrickerBot - NJCCIC Threat Profile*. [online] Cyber.nj.gov. Available at: <https://www.cyber.nj.gov/threat-center/threat-profiles/botnet-variants/brickerbot> [Accessed 3 May 2021].

- OWASP, 2018. *OWASP IoT Top 10*. [online] Owasp.org. Available at: <https://owasp.org/www-pdf-archive/OWASP-IoT-Top-10-2018-final.pdf> [Accessed 17 May 2021].

- Babar, S., Mahalle, P., Stango, A., Prasad, N. and Prasad, R., 2010. Proposed Security Model and Threat Taxonomy for the Internet of Things (IoT). In: *International Conference on Network Security and Applications*. [online] Chennai: Springer, pp.420-429. Available at: <https://link.springer.com/chapter/10.1007/978-3-642-14478-3_42> [Accessed 1 May 2021].

- Pacheco, J. and Hariri, S., 2016. IoT Security Framework for Smart Cyber Infrastructures. In: *1st International Workshops on Foundations and Applications of Self\* Systems*. [online] Augsberg: IEEE, pp.242-247. Available at: <https://ieeexplore.ieee.org/abstract/document/7789475> [Accessed 1 May 2021].

- Alhabi, R. and Aspinall, D., 2018. *An IoT Analysis Framework: An Investigation of IoT Smart Cameras' Vulnerabilities*. Edinburgh: University of Edinburgh, pp.1-10.

- Sanford, M., Woodraska, D. and Xu, D., 2011. Security Analysis of FileZilla Server Using Threat Models. In: *Proceedings of the 23rd International Conference on Software Engineering & Knowledge Engineering*. Miami Beach: ResearchGate, pp.678 - 682.

- Xiong, W. and Lagerström, R., 2019. Threat Modeling of Connected Vehicles: A privacy analysis and extension of vehicleLang. In: *International Conference on Cyber Situational Awareness, Data Analytics and Assessment*. Oxford: IEEE, pp.1-7.

# 10. Appendices

## 10.1 Appendix A - State of the Market for Threat Modelling Tools

| Name | Pros | Cons | Price | Link |
|------|------|------|-------|------|
| ThreatModeler | Continuously updated with new research findings, advanced threat prediction patented tools, comes with network template tools, cloud-specific 'automated threat library' | Intended only for corporate user, inaccessible to the beginner or novice home user, very expensive, requires high level of background knowledge | Free 30 day demo, $4000 dollars for the shortest license (1yr) | https://threatmodeler.com/ |
| Forward Networks | Verifies and predicts network security through modelling a 'twin', usage endorsed by well-known companies, easy to search for and find a specific problematic device on a very large network, model proposed changes before enacting them | Requires high level of technical knowledge, intended only for corporate use | Contact for quote | https://forwardnetworks.com/ |
| ENISA Good Practices for IoT and Smart Infrastructures Tool | Provides an exhaustive checklist of the desirable security measures, can filter by security measure type, accessible in browser, includes references to further reading under each measure, filter by application (base IoT, smart cars, hospitals etc) | Does not allow modelling of a specific network, heavy use of technical jargon impying expertise | Free | https://www.enisa.europa.eu/topics/iot-and-smart-infrastructures/iot/good-practices-for-iot-and-smart-infrastructures-tool/results#IoT |
| ITarian Network Assessment Tool | Identifies vulnerabilities across the network, produces automatic reports, setup wizard, monitoring tools for adding new devices, identify where upgrades are needed, generates SWOT analysis | Intended for experienced administrators, intended for corporate use, complex interface | Free for up to 50 devices | https://itarian.com/app/ |
| N-Able | Calculates monetary costs of security failures, discovery and alert dynamic system, generates reports | For corporate use, intended for experienced professionals, based primarily around financial | Contact for quote | https://www.n-able.com/features/risk-assess |
| | | implications for businesses such as lawsuits for data loss | | ment-softw are-tools#ri sk-assessm ent |
| SolarWinds Network Assesment | Maintains up-to-date network inventory, automatic device discovery, highlights violations from vulnerability scanning, offers network configuration backup and restore, automated auditing | High level of expertise required, intended for corporate use, expensive | Free trial, after that starts around £1300 | https://ww w.solarwin ds.com/net work-confi guration-m anager |
| Lansweeper | Discovery of all assets on network, creates audits that highlight risks and mitigation methods, very cheap for small scale use, regular updates, event logs | Requires high level of technical knowledge, intended for corporate use | Free for up to 100 assets, after that $1 per asset per year | https://ww w.lansweep er.com/ |
| Nmap | Maps network and all active connections, free forever, extremely powerful tools | Does not provide recommendations or reports, requires very high level of prior expertise, intended for security experts, runs in terminal unless using less powerful Zenmap | Free | https://nma p.org/ |

## 10.2 Appendix B - Rough Concept Designs

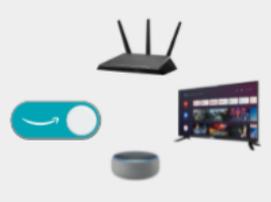

## 10.3 Appendix C - Supervisory Meeting Log

| Meeting 1 |
| --- |
| *Agenda: Discuss chosen methodology and chosen IoT scenario. Discuss preliminary research steps taken. Ask about primary research proposal and ethical approval.* |
| <ul><li>There is no restriction on how many types of home assistant that can be covered in the scenario, but it is probably best to start with or focus on 1 for the sake of feasibility.</li><li>Apply multiple threat model techniques and compare them</li><li>The more new knowledge you are creating, the better</li><li>RE: Ethical approval, speak with Dr Wong</li></ul> |
| *Next steps: Condense new ideas and papers of interest and send these over [completed].* |

| Meeting 2 |
| --- |
| *Agenda: Discuss conversation with Sylvia and talk about Helsinki seminar attendance. Go over the structural skeleton plan and plot next steps.* |
| <ul><li>Number the sections in the report</li><li>Create a project management timeline to be contained in appendices</li></ul> |

- In the conclusion, reflect on skills and learning, including potential limitations
- Both a bibliography and reference list are not wholly necessary as long as you have one or the other

*Next steps: Create a meeting log [completed]. Draft a research design plan for the ethical proposal [completed]. Create a rough timeline of the project [completed].*

---

**Meeting 3**

*Agenda: Discuss feedback from Dr Wong, and the project timeline (which may now be subject to some changes).*

- A software development segment is required. Depending on the scope of the accompanying research, this software does not have to be huge.
- The full ethical guideline design is not due until mid-June. This gives some leeway time for finalising the full timeline of the project, including a small development segment.
- It is important to have an approved software development idea first before undertaking any requirements analysis steps.

*Next steps: Meet with Sylvia to discuss the software idea. Be sure to keep in touch to discuss next steps following an approved software idea.*

---

**Meeting 4**

*Agenda: Discuss software development idea*

- The software will allow the user to select devices in their home IoT environment, as well as how these devices are connected. It will then show the user the home environment layout they have selected and detail potential risks, along with how to mitigate them.
- The software might link the user to manufacturer guidance for their devices.
- In the back end, threat modelling techniques will be applied to rank the threats by severity when they are displayed to the user.
- Limited number of devices to start as this is a proof of concept.

*Next steps: Create some mockup screen sketches to illustrate the idea [completed].*

---

**Meeting 5 - With Dr Wong**

*Agenda: Discuss software development proposal and gauge next steps for preliminary document submissions*

- The idea will be good and can be tested with participants, which will require some extra ethical consideration.
- The software idea will be useful, though some diagnostic feature (e.g. checkbox questions) could also be added to provide a more personalised view of the user's environment.
- Could be done client-side (e.g. in HTML) or server side, which would allow collection of data for future research, perhaps? Ultimately it is down to how much time is available.
- Consider the types of questions the research project is aiming to answer overall when drafting requirements.
- Adding more information to the preliminary documents would potentially help in planning.
- The project will mainly have academic and social benefits.
- As long as it's not a waterfall development approach, it's fine!

*Next steps: Elaborate on the preliminary documents to reflect new steps and concepts. Conduct a more intricate requirements analysis nearer the time, once initial survey data elucidates the problem.*

---

**Meeting 6**

*Agenda: Present ideas for the main questions the project will answer. Discuss the developing ideas for software, including how it will connect to the theory.*

- Research questions proposed make sense. Now plan out the bullet point steps of how these questions will be tackled in the project
- Doing this will make it more clear to people how the research will link to the produced software
- A plan for how to answer each research question will eventually go in the introduction of the paper
- Next time may discuss how to start tackling all the theory analysis

*Next steps: Bullet point plan of how each question will be answered. Email update later in the week. May receive some midweek feedback on the content of the newer drafted project proposal, so keep an eye out for any emails.*

**Meeting 7**

*Agenda: Discuss what has been completed in the last week and consider the next ethical submissions. Talk about how to start making threat models.*

- Not needed to have so many data collection steps
- Focus on software feedback as primary data gain
- Base requirements analysis on prior work and propose a new requirement for own software; it will take too long if users drive requirements
- Take a look at other threat modelling software, how will yours be unique?

*Next steps: Aim to complete ethical stuff with new ideas by Monday. Email midweek. Document current research into modeling approaches for the sake of proposal form.*

**Meeting 8**

*Agenda: Finalize all documentation for submission today. Discuss the next step i.e. modelling approaches.*

- Paperwork is looking okay, be on the lookout for an email later giving the all clear
- For model comparison, start with 3 models and aim for 5
- STRIDE, CVSS, Trees, LINDUNN (GO), Quantitative
- Consider pros and cons of the modelling approaches, relating to scenario chosen
- Will create a table to evaluate them on different metrics
- When it comes to software requirements analysis, report back for help

*Next steps: Be on the lookout for the OK email. Start looking at threat modelling.*

**Meeting 9**

*Agenda: Discuss progress this week (diagrams and models). Talk about next ethics doc deadline.*

- Can use student groups internally for recruitment if needed
- Send over diagrams when computerized
- It's okay to not do other ones like PASTA, but explain why they're not feasible here
- Form ready to submit

*Next steps: Submit form today. Send over diagrams this week.*

**Meeting 10**

*Agenda: Discuss progress this week (last threat modelling bits). Talk about next steps (researching software that already exists).*

- Two minor edits for the scenario diagram. Specify that the 'hub' is a VA hub, and add a second arrow for connected devices and users.
- Next step is to research different software that is already out there, perhaps arranging

in a matrix. Pros, cons, gaps?
- In the final paper, software requirements analysis could come first, and then slot the threat modelling inside that
- Use agile approach for software development, with 1-2 week sprints
- A MoSCoW table may help for planning software

*Next steps: Researching software. Look into MoSCoW method.*

---

**Meeting 11**

*Agenda: Discuss what has been done in the past week and what must be done next.*

- Looked into some preliminary info about req analysis
- Pros and cons matrix for threat models
- Research into existing state of threat modelling software

*Next steps: Begin req analysis process*

---

**Meeting 12**

*Agenda: Discuss requirements analysis document and current sprint progress*

- Requirements analysis looks fine.
- Coding progress going well.
- Make sure to specify in the write up that coding comes second and the primary focus of the thesis is the analysis work
- Keep a log of what's done per sprint and what's left to do next sprint

*Next steps: Annotate mock-up screenshots with what the user will do on each page. Continue coding work.*

---

**Meeting 13**

*Agenda: Progress update.*

- Progress is going to plan
- After MVP is done, start writing up proper analysis in parallel with 'bonus' features and tweaks
- MVP is most important thing at this point.

*Next steps: Keep working to sprint schedule.*

---

**Meeting 14**

*Agenda: Discuss progress on coding.*

- Coding for MVP is done
- Added some additional features on top of MVP
- Have some potential users for testing, ethical approval is ready

*Next steps: Finish tweaks and begin user testing*

---

**Meeting 15**

*Agenda: Talk about progress with testing and when to begin write-up*

- User testing has begun and will continue for the rest of the week
- Ensure proper documentation trail for each user test
- Consent forms need to be saved for thesis defense but do not need to be appended to the dissertation document itself
- Next week begin writing up section by section, send over potential completed sections separately for review

*Next steps: Keep working to schedule*

**Meeting 16**

*Agenda: Discuss write-up progress and final structure*

- Abstract and acknowledgements and contents don't need section number, intro starts from 1
- 'This was the form used for taking the feedback' - not mandatory but maybe
- Keep bibliography in after reference list now it is complete

*Next steps: Wait for further feedback*

*10.4 Appendix D - Algorithm Documentation*

`Algorithm.md`

```
#THE ALGORITHM - HOW IT WORKS

    Copyright (C) 2021  Beckett LeClair

    This program is free software: you can
redistribute it and/or modify
    it under the terms of the GNU General Public
License as published by
    the Free Software Foundation, either version 3 of
the License, or
    (at your option) any later version.

    This program is distributed in the hope that it
will be useful,
    but WITHOUT ANY WARRANTY; without even the
implied warranty of
    MERCHANTABILITY or FITNESS FOR A PARTICULAR
PURPOSE.  See the
    GNU General Public License for more details.

    You should have received a copy of the GNU
General Public License
    along with this program.  If not, see
<https://www.gnu.org/licenses/>.

This documentation is intended to explain and justify
the design of the calculator in the software. It is
intended for those with strong background knowledge
of threat modelling.
```

The algorithm takes elements from STRIDE, CVSS, and LINDUNN GO.

Firstly, the devices the user selects are sorted into the following five categories. A device can be in more than one category at once. If at least one device per category is added, that category is switched on for the calculations.
- Voice input devices (Cat1)
- Devices generally requiring sign-in to function (2)
- Devices communicating on the internal network (3)
- Devices communicating with external entities via the internet (4)
- Devices directly affecting home physical security (5)

The key to device categories is as follows:
- Home virtual assistant (1 2 3 4 5)
- Smart security cam (2 3 4 5)
- Smart doorbell (2 3 5)
- Smart lighting (3)
- Smart fitness aid (2 3 4)
- Smart kitchenware (3)
- Smart locks (3 4 5)
- Amazon Dash (2 3 4)
- Smart thermostat or air monitor (2 3 4 5)
- Automated 'smart home' controller* (1 2 3 4 5)
- Smart sleep tracker (2 3)
- Any other smart home devices** (1 2 3 4 5)

Obviously, these are just averages or assumptions based on what most devices of these types do. In a more developed version of the program, individual makes/models and their unique capabilities could be taken into consideration.

1* These fulfil generally the same exact functions as the virtual assistant hub and often have touchscreens as well.
2** We will have to assume these devices in the 'other' class are reasonably capable of falling into any category.

As the different devices follow differing amounts of the same threats already elucidated for the voice processing hub acting as central controller (see research paper), the same STRIDE table can be used (albeit a filtered down version by device type for relevancy) when determining the key risks.

The switched-on categories will then turn on the associated threat numbers. These numbers correlate to the STRIDE table (see paper).

- Cat 1: S(1), T(4 7), I(10 11), D(12)
- Cat 2: S(2), I(8)
- Cat 3: S(3), T(6 7), I(9), D(13 14), E(15 16)
- Cat 4: S(2), T(5), I(8 11), D(13), E(15)
- Cat 5: S(3), T(6 7), I(9), D(14), E(16)

The risk score added per STRIDE number is the average of the 3 different CVSS scores (see paper) for that threat.

Now, the risk factors toggled by the user will also add a value to the score, based on how easy it might be for a stranger to manipulate the risk factor to form an attack. These scores go from 1 (the attacker would have a hard time utilizing this specific route though it is still possible) to 3 (relatively easy to manipulate). Obviously there is a degree of opinion to these values, but it would be very simple to change them in the code to better suit future developments in knowledge. The STRIDE values match to their corresponding calc variable.
- R1: value 3, related STRIDEs 2 3 5 6 7 8 9 12 13 14 15 16, justification: it is incredibly easy to snoop on traffic in an unprotected network.
- R2: value 1, related STRIDEs 1 6, justification: this removes a step of complexity in gaining account access although the attacker must still manage to crack the first passcode in most cases
- R3: value 2, related STRIDEs 2 3 5 6 7 8 9 12 13 14 15 16, justification: the default passwords on many models of router can be brute forced with little effort but this depends on the victim having certain models of router
- R4: value 2, related STRIDEs 2 3 6 9 15 16, justification: weak passwords are easily brute-forced but doing this still requires significant processing power in many cases
- R5: value 1, related STRIDEs 1 4 10 12, justification: manipulating this would require an attacker to position themselves very close to the victim at an opportune moment
- R6: value 1, related STRIDEs 11, justification: there are multiple documented cases of employees hearing private conversations through this processing though obviously it depends on a malicious employee happening to hear a specific thing
- R7: value 2, related STRIDEs 1 11 12, justification: a malicious skill/app can turn on the listening mode but this requires the user to have the skill/app in the first place or for the attacker to be able to escalate privileges high enough to toggle the mic
- R8: value 3, related STRIDEs 15 16, justification:

it is very easy to search for device vulnerabilities online and in many cases example attack scripts are provided
- R9: value 3, related STRIDEs 2 3 5 6 7 8 11 13 14 15 16, justification: once a malicious app is installed then an attacker has very easy access to the device they want to manipulate
- R10: value 3, related STRIDEs 2 3 5 6 7 8 9 11 13 14 15 16, justification: as above, but for devices instead of apps
- R11: value 2, related STRIDEs 1 10 11 12, justification: it would require a bad actor to be very close though many accidental bad actors manipulate this unknowingly or as a joke
- R12: value 2, related STRIDEs 9, justification: based on usage traffic patterns (even encrypted ones) an attacker can easily find out when someone is not home and use that information maliciously
- R13: value 1, related STRIDEs 11, justification: this risk relies on the third party having malicious actors within it
- R14: value 1, related STRIDEs 11, justification: this relies on a manufacturer being malicious

Now, we can take away 1 from calc scores for each potential mitigation for the threat is in place (by virtue of the opposite risk factor not being turned on). A threat value (calc) is not allowed to go below 0:
- If R1 is not on, reduce the threat values for fake server signals and fake device signals.
- If R2 is not on, reduce the threat value for outsider commands.
- If R3 is not on, reduce the threat values for personal data leaks, compromised server signals, compromised action signals, and action leaks.
- If R4 is not on, reduce the threat values for personal data leaks, compromised action signals, and action leaks.
- If R5 is not on, reduce the threat values for eavesdroppers and interfering commands.
- If R6 is not on, reduce the threat value for private conversation leaks.
- If R8 is not on, reduce the threat values for compromised server signals and compromised action signals.
- If R9 is not on, reduce the threat values for congesting server signals and congesting action signals.
- If R10 is not on, reduce the threat values for congesting server signals and congesting action signals.
- If R12 is not on, reduce the threat value for

action leaks.
- If R13 is not on, reduce the threat value for
private conversation leaks.

Each category also has a number of POTENTIAL
associated LINDUNN GO factors (again, individual
devices would vary, but it is not feasible to analyse
every brand and product for this project).
- Cat 1 has 11 factors: linkable user actions,
linkability of retrieved data, actions identifying
user, non-repudiation of sending, non-repudiation of
received data, no transparency, disproportionate
collection, unlawful processing, disproportionate
processing, identifiability from shared data,
identifiability from inbound data.
- Cat 2 has 17 factors: linkable user credentials,
linkable user actions, linkable inbound data,
linkable shared data, linkable stored data, linkable
retrieved data, identifying credentials, actions
identifying user, identifying inbound data,
identifying shared data, identifying retrieved data,
non-repudiation of sending, non-repudiation of
storage, non-repudiation of retrieval,
disproportionate collection, unlawful processing,
disproportionate storage.
- Cat 3 has 4 factors: linkable credentials, linkable
context, identifiable context, detectable outliers.
- Cat 4 has 10 factors: linkable inbound data,
linkable shared data, linkable retrieved data,
identifying shared data, identifying retrieved data,
detectable outliers, disproportionate collection,
unlawful processing, disproportionate processing,
disproportionate storage.
- Cat5 has 4 factors: linkable user actions, linkable
context, identifying context, detectable outliers.
A score is added to each risk in the category based
on the number of LINDUNN GO categories flagged by the
device type, multiplied by 0.5. This 0.5 was chosen
as some device types have a very large amount of
associated GO factors compared to others, so dividing
the total by 2 reduces overly-skewed results.

Finally, if we have covered both mitigations for
private conversation leaks (shown by R6 and 13 both
being unselected), we can remove calc11 by setting it
to 0 as it is no longer a problem.

The remaining results will be sorted and have any 0
results removed so they can be output to the user in
descending order (biggest threats first).

README.txt

*My Threat Model*

*To start the application, run the supplied executable file.*

*This application is designed to let home users learn more about the threat environment facing their IoT devices. It allows users to model their environment by selecting devices and risk factors, and then ranks threats according to the level of risks, also giving more details about why each one is a risk and some potential ways it could be mitigated. Scores are dimensionless and are calculated based on user inputs, giving a relative point of comparison for severity.*

*This is just a proof-of-concept. The idea is for this application to be used as a starting point, so others can work on making a threat modelling software that is accessible to the average consumer. The information presented in some sections assumes that the user is based within the UK, though the general advice is still applicable to all.*

*This software is a novel concept as there is a distinct lack of any threat modelling software that is a) intended for the average consumer who is not already an expert in infosec, and b) available at a low cost (free in this case).*

```
    You should have received a copy of the GNU
General Public License
    along with this program.  If not, see
<https://www.gnu.org/licenses/>.
```

## 10.6 Appendix F -Ethical Risk Assessment

| Risks concerning data collection from the general public | | | |
|---|---|---|---|
| **Potential Risk** | **Risk Level** | **Justification** | **Mitigation** |
| Lone working with members of the public which may lead to risk of harm | Low | Research will not involve physical contact | Remote research methods such as Microsoft Teams negate need for physical contact |
| Potential for inappropriate or abusive data responses | Medium | Could possibly occur, but are easily filtered out and/or discarded | This risk must be accepted due to the nature of the project |
| Insecure data storage leading to violation of privacy legislation and best practice | Low | Only trusted, GDPR-compliant devices and encrypted file storage services to be used | Data will be stored in a private drive protected by 2FA and a strong password known only by one person; laptop containing research files will be in a secured location and will not leave the UK; all data will be deleted post-release of a final dissertation grade |
| Violation of respondent anonymity | Low | No identifying data will be stored long-term | Personally identifying data will not be collected; respondents will be assigned number codes for anonymity; audio will be deleted immediately following transcript creation, no contact details will be stored long-term |
| Some respondents might not want to disclose experience regarding their personal feelings of trust and device security | Low | Nature of content is very unlikely to be upsetting | Respondents are to be assured that taking part is completely voluntary and they can back out at any time without being penalised |

| Risks concerning software development and other theoretical aspects of the project | | | |
|---|---|---|---|
| **Potential Risk** | **Risk Level** | **Justification** | **Mitigation** |
| May not be sufficient time to realise all potential software ideas | Medium | Potential to happen, but mitigating measures are in place | Detailed time planning will discourage likelihood of falling behind; items assigned priority levels so the most important work will always be done first; RAD approach |
| Insecure storage leading to data loss | Low | Due to data storage methods, loss is unlikely | Work in progress will largely be stored on a password-protected cloud drive with automatic saving and recovery |
| Software errors may cause unwanted effects on testers' devices | Low | Software will be heavily prepared before this stage | Software will be pre-tested on several devices belonging to the researcher to prevent compatibility problems; code will be screened for flaws multiple times before user testing |